\documentstyle{mn}

\input epsf.tex

\def\fm{\hbox{$.\!\!^{\rm m}$}}
\def\farcs{\hbox{$.\!\!^{\prime\prime}$}}
\def\kms{\rm km\,s^{-1}}
\def\Ie{{< \hspace{-3pt} I \hspace{-3pt}>_{\rm e}}}
\def\re{{r_{\rm e}}}
\def\Mgtwo{{ {\rm Mg}_2}}
\def\Mgone{{ {\rm Mg}_1}}
\def\Fe{{ <\hspace{-1.5pt}{\rm Fe}\hspace{-1.5pt}>}}
\def\Hb{{ {\rm H}{\beta}}}
\def\HbG{{ {\rm H}{\beta}_{\rm G}}}
\def\MrT{{M_{\rm r_T}}}
\def\Ho50{{H_{\rm o}=50~{\rm km\, s^{-1}\, Mpc^{-1}} }}

\title[Stellar populations of cluster E and S0 galaxies]
{Stellar populations of cluster E and S0 galaxies}
\author[I. J{\o}rgensen]
{Inger J{\o}rgensen\thanks{
E-mail: inger@roeskva.as.utexas.edu
\newline $\dagger$ Hubble Fellow. }$^{\,\dagger\, }$  \\
McDonald Observatory, The University of Texas at Austin, RLM 15.308, Austin, TX 78712, USA }
\date{8 February 1997. Accepted for publication in MNRAS.}

\begin{document}

\maketitle

\begin{abstract}

Spectral line index data for a sample of 290 E and S0 galaxies are
used to investigate the stellar populations of these galaxies.
250 of the galaxies are members of 11 nearby clusters 
($cz_{\rm CMB} < 11500\kms$).

It is studied how the stellar populations of the galaxies are related 
to the velocity dispersions, the masses of the galaxies, and the 
cluster environment.  This is done by establishing relations between 
these parameters and the line indices $\Mgtwo$, $\Fe$ and $\HbG$.
The difference between the slope of the $\Mgtwo$-$\sigma$ relation
and the slope of the $\Fe$-$\sigma$ relation indicates that 
the abundance ratio
[Mg/Fe] is 0.3-0.4 dex higher for galaxies with velocity dispersions
of 250$\kms$ compared to galaxies with velocity dispersions of 
100$\kms$. This is in agreement with previous estimates by Worthey et al.

The $\Fe$ index is stronger correlated with the projected cluster 
surface density, $\rho _{\rm cluster}$, than with
the galaxy mass or the velocity dispersion. We have earlier found the
residuals for the $\Mgtwo$-$\sigma$ relation to depend on the
cluster environment. Here we determine how both the $\Mgtwo$ index and 
the $\Fe$ index depend on the velocity disperson 
and $\rho _{\rm cluster}$. Alternative explanations that could create 
a spurious environment dependence are discussed. No obvious 
alternatives are found.  The environment dependence of 
the $\Mgtwo$-$\sigma$ relation is supported by data from Faber et al.
The dependence on the environment implies that [Mg/Fe] decreases
with increasing density, $\rho _{\rm cluster}$.
The decrease in [Mg/Fe] is 0.1 dex over 2.5 dex in $\rho _{\rm cluster}$.

We have also studied to what extend the mass-to-light (M/L) ratios of 
the galaxies are determined by the stellar populations.
The M/L ratios are strongly correlated with the indices $\Mgtwo$ and 
$\HbG$, while the $\Fe$ index is only weakly correlated with the M/L ratio.

Based on current stellar population models we find that it is not
yet possible to derive unique physical parameters (mean age, 
mean abundances, mean IMF, and fraction of dark matter) from the 
observables (line indices, velocity dispersion, mass, M/L ratio).

\end{abstract}


\section{Introduction}

Optical studies of stellar populations in external galaxies beyond the 
Virgo cluster are limited to investigations based on the integrated 
light of the stars in smaller or larger parts of the galaxies.
Radial gradients can be studied, but the individual stars cannot
be resolved with present day instrumentation.  The interpretation of the
observations therefore relies on models of the stellar populations.

Since Baade (1944) introduced the idea of stellar populations in 
external galaxies broad band colors have been used for observational 
studies as well as modeling of stellar populations
(e.g., Tinsley \& Gunn 1976; Aaronson et al.\ 1978;
Bruzual 1983; Buzzoni 1989, 1995; Bruzual \& Charlot 1993;
Peletier \& Balcells 1996).
More detailed studies are possible based on measurements of absorption 
features. In the Lick/IDS system indices are defined for lines in the
optical region, 4100--6300 {\AA} (Faber et al.\ 1985;
Worthey et al.\ 1994).
Line strengths for galaxies have been measured in this system by, 
e.g., Gorgas, Efstathiou \& Arag\'{o}n-Salamanca (1990),
Worthey, Faber \& Gonz\'{a}les (1992), Davidge (1992),
Davies, Sadler \& Peletier (1993), Carollo, Danziger \& Buson (1993),
Fisher, Franx \& Illingworth (1995, 1996) and Vazdekis et al.\ (1996ab).

Recent stellar population models have made predictions
of the line indices in the Lick/IDS system, 
the broad band colors and the mass-to-light (M/L) ratio
for single stellar populations (Buzzoni, Gariboldi \& Mantegazza 1992;
Worthey 1994; Buzzoni, Mantegazza \& Gariboldi 1994;
Vazdekis et al.\ 1996a; Bressan, Chiosi \& Tantalo 1996).
These models are static models in the sense that the aim is not to 
model the evolution of a stellar system, but to predict the observables 
for one stellar population with a given age, metallicity and initial 
mass function (IMF).  Worthey and Bressan et al.\ use a Salpeter (1955)
IMF, while Buzzoni et al.\ and Vazdekis et al.\ present models for 
several IMFs.  Most of the available models use solar abundance ratios,
specifically [Mg/Fe]=0. 
Current static models with [Mg/Fe]=0 fail to reproduce the observed 
flat relation between the magnesium index $\Mgtwo$ and the iron 
index $\Fe$ found for elliptical galaxies (Worthey et al.\ 1992; 
Buzzoni et al.\ 1994).  Weiss et al.\ (1995) studied the effect of a 
non-solar [Mg/Fe] and estimated that bright elliptical galaxies have
[Mg/Fe] in the interval +0.3 to +0.7.

Some authors have investigated the possibility of fitting
evolutionary models to the stellar populations in elliptical
galaxies. Even though beyond the scope of this paper,
we note that the main problem is to create sufficiently enriched and
old populations of stars that will fit the red colors and
strong line indices characteristic for elliptical galaxies.
Vazdekis et al.\ (1996ab) find that some process
is needed to supply a strong enrichment early in the evolution
of the galaxies. They argue that a time variable IMF may be feasible.
Bressan et al.\ (1996) use an infall model to fit line index
data from Gonz\'{a}les (1993) and UV data from Burstein et al.\ (1988a).

Models of single stellar populations can be used to interpret the 
measured line indices and broad band colors in the sense that the
``best fitting'' model gives an estimate of the luminosity weighted mean
age, mean metallicity and mean IMF of the current stellar populations.
One of the problems with this technique is the degeneracy 
of the effects from variations in age and in metallicity.
In general broad band colors cannot be used to disentangle
age and metallicity effects (Aaronson et al.\ 1978; Worthey 1994).
The possible presence of dust in E and S0 galaxies adds further 
confusion to the interpretation of broad band colors (e.g., 
Silva \& Elston 1994).  Many of the spectral indices defined in the 
Lick/IDS system suffer from the same degeneracy regarding age and 
metallicity. Worthey (1994) argues that $\Hb$ together with one of the 
magnesium indices can be used to break the degeneracy, since $\Hb$
is more sensitive to age than the magnesium indices.
In later studies Jones \& Worthey (1995) and 
Worthey, Trager \& Faber (1995) use higher order Balmer lines together
with several indices for heavier elements to estimate
mean ages and metallicities for a sample of E galaxies.
The current models are not in agreement with regard to the 
predicted near-infrared broad band colors.
Peletier \& Balcells (1996) find that near-infrared colors to some 
extent can be used to solve the age/metallicity degeneracy problem for 
E galaxies, while Worthey (1994) does not find near-infrared colors 
useful for breaking the degeneracy.
The main cause of the differences in the near-infrared colors
is due to differences in the adopted models for the stellar evolution
(Charlot, Worthey \& Bressan 1996).
None of the techniques for breaking the age/metallicity degeneracy
takes variations in the abundance ratio [Mg/Fe] into account.

Ideally one wants to use the observables
(the line indices, the velocity dispersion, the M/L ratio, etc.)
to derive physical parameters like the mean age, the mean
abundances of various heavy elements, and the mean IMF for the stellar 
populations presently observed. In this context it should also be
addressed what the fraction of dark matter is in the galaxies. 
If the fraction of dark matter is not the same in all E and S0 galaxies,
this may give variations in the M/L 
ratio that are not reflected in the observed line indices.

In this paper spectroscopic data are analyzed for a large sample of 
cluster E and S0 galaxies. We concentrate on centrally measured indices
aperture corrected to a standard size aperture. The aim is to establish
empirical relations between the observables (the velocity dispersion; 
the line indices $\Mgtwo$, $\Hb$ and $\Fe$; the M/L ratio; and the mass). 
The underlying physical questions are (a) whether the mix of stellar 
populations is determined by the velocity dispersion (or alternatively 
the mass), (b) which influence the cluster environment has,
and (c) if the M/L ratio is determined by the stellar populations.
The data are also compared to predictions from models of single 
stellar populations and the expected variations in the observables
due to changes in the age, the abundances, the IMF, and 
the fraction of dark matter are discussed.

There are several studies of the $\Mgtwo$-$\sigma$ relation for large 
samples of E and S0 galaxies (Burstein et al.\ 1988b; 
Guzm\'{a}n et al.\ 1992; Bender, Burstein \& Faber 1993; 
J{\o}rgensen, Franx \& Kj{\ae}rgaard 1996, hereafter Paper II).
Relations that also involve $\Hb$ and $\Fe$ have previously
only been studied for relatively small samples of galaxies (e.g.,
Gorgas et al.\ 1990;
Worthey et al.\ 1992; Davies et al.\ 1993; Gonz\'{a}les 1993; 
Worthey 1994; Fisher et al.\ 1995, 1996), though relations for a larger 
sample of E galaxies are shown by Burstein et al.\ (1984).

The present paper is organized as follows. Sect.\ 2 briefly describes 
the samples of galaxies.  The determination of the line indices is 
covered by Sect.\ 3 and the Appendix.  Sect.\ 4 has the empirical point
of view. Linear relations between the available observables are 
established.  It is also tested how the relations depend on the 
cluster environment.  In Sect.\ 5 the data are compared with model 
predictions from stellar population models. 
The conclusions are summarized and discussed in Sect.\ 6.

Unless otherwise noted the relations established in this paper are 
determined by minimization of the sum of the absolute residuals 
perpendicular to the relations, and the zero points are derived as the 
median zero points.  This fitting technique has the advantage that it 
is rather insensitive to a few outliers, and that it treats the
coordinates in a symmetric way.
The uncertainties of the coefficients are derived by a bootstrap
procedure. See also Paper II for a discussion of this fitting technique.

\begin{table}
\caption[]{Number of E and S0 galaxies with available line indices 
\label{tab-cluster} }
\begin{tabular}{lrrr}
Cluster    & $\HbG$ &  $\Mgtwo ^{\rm a}$ & $\Fe$ \\ \hline
Coma       &    & 80 &    \\
Abell 194  & 19 & 19 & 19 \\
Abell 539  &  1 & 29 & 29 \\
Abell 3381 & 16 & 16 & 16 \\
Abell 3574 &  7 &  7 &  7 \\
Abell S639 &  4 & 21 & 21 \\   % omit J31
Abell S753 & 14 & 14 & 14 \\
DC2345-28  &    & 10 &    \\
Doradus    &  8 &  8 &  8 \\
HydraI     & 12 & 42 & 42 \\
Grm15      &  4 &  4 &  4 \\ \hline
Cluster sample    & 85 &250 &160 \\ 
Additional sample & 40 & 40 & 39 \\ \hline
\end{tabular}

Note -- $^{\rm a}$ Literature data for $\Mgtwo$ included, cf.\ Sect.\ 3.\\
$\Mgtwo$ values derived from Mgb are also included. \\
$\Fe$=(Fe5270+Fe5335)/2.
\end{table}

\section{Sample selection and data}

The observational data were originally obtained for our study of the 
Fundamental Plane for E and S0 galaxies in 11 nearby clusters 
(Paper II).  The selection criteria for the galaxies are described in 
detail in J{\o}rgensen, Franx \&
Kj{\ae}rgaard (1995a) and J{\o}rgensen \& Franx  (1994).
The main selection criteria were classification (E or S0) 
and total magnitude. It should be noted that the samples are
not complete to a well-defined absolute total magnitude.
A representative lower limit on the luminosity of the galaxies
in the sample is $\MrT = -20\fm 45$ in Gunn r ($\Ho50$).

Photometry in Gunn r is taken from J{\o}rgensen, Franx \& Kj{\ae}rgaard
(1995a).  The velocity dispersions and the $\Mgtwo$ indices are from
J{\o}rgensen, Franx \& Kj{\ae}rgaard
(1995b, hereafter Paper I) and the literature, see Sect.\ 3.
In Sect.\ 3 other line indices are derived from the same spectroscopic
observations used in Paper I.

From the available data we define two samples.
The ``cluster sample'' consists of the 250 E and S0 galaxies
which are members of the 11 nearby clusters and for which we have at 
least the velocity dispersion and the $\Mgtwo$ index, 
see Table \ref{tab-cluster}.
Information about the cluster properties (radial velocity,
richness, cluster velocity dispersion, etc.) is given in Paper II.
207 of the galaxies also have reliable photometry.
(Though photometry is available for the galaxies Coma-D120 and 
Coma-D121 it is not used here, because the small angular distance 
between the two galaxies makes the photometric parameters highly 
uncertain.) The ``additional sample'' consists of the E and S0 galaxies
observed for comparison purposes, some E and S0 galaxies in the 
Hickson (1982) compact groups, and observed E and S0 galaxies that 
turned out not to be members of the 11 nearby clusters.
There are 40 galaxies in the additional sample.
The cluster environments for these galaxies are in general
less dense than for the galaxies in the cluster sample
as most of the galaxies belong to small groups or the field.

\section{Spectroscopic data}

The spectra were obtained between 1990 and 1992 during three observing 
runs at the ESO 1.5m telescope equipped with the Boller \& Chievens
spectrograph (hereafter B\&C) and one observing run at the ESO 3.6m 
telescope with the OPTOPUS instrument, a fiber-fed B\&C spectrograph.
Full information about the observing runs and the instrumentation
is given in Paper I.
The B\&C spectra covered the wavelength interval 4700-5700{\AA} or
4450-6160\AA. The coverage for the OPTOPUS spectra was 5000-5620\AA.
The instrumental resolution was typically 1.25{\AA} measured as
sigma in a Gaussian fit to lines in a calibration spectrum.
The typical signal-to-noise (S/N) ratio per {\AA}ngstr\"{o}m
is between 20 and 40.
In total 220 galaxies were observed.
Paper I describes the basic reduction of the spectra and the 
determination of radial velocities, velocity dispersions,
and the $\Mgtwo$ index.
The typical uncertainties derived from external comparisons are
$\pm 0.036$ for $\log \sigma$ and $\pm 0.013$ for $\Mgtwo$, see Paper I.
Here we determine the indices for additional absorption lines.
The indices are derived from the flux calibrated spectra.
We adopt the Lick/IDS definitions for the line indices (Worthey 
et al.\ 1994).  The C4668 index is called Fe4668 by Worthey et al.
We prefer to refer to it as C4668, because it is highly sensitive
to the carbon abundance and not to the iron abundance 
(Tripicco \& Bell 1995).
Table \ref{tab-index} lists which indices have been measured and 
the number of galaxies for which it was possible to measure each index.
The observations of S639-J31 and DC2345-28-D38 had too low S/N ratios
to derive useful line indices. The median internal uncertainties 
based on the S/N ratio of the spectra are also given in the table.

\begin{table}
\caption[]{Number of measured indices and their median internal uncertainties
\label{tab-index} }
\begin{tabular}{lrrr}
Index         & N$_{\rm gal}$ & $\sigma _{\rm index}$ 
              & $\sigma _{\rm log (index)}$ \\ \hline
Fe4531   &  27 & 0.46 & 0.072 \\
C4668   &  58 & 0.70 & 0.043 \\
$\Hb$    & 135 & 0.26 & 0.064 \\
$\HbG$   & 131 & 0.17 & 0.038 \\
Fe5015   & 213 & 0.47 & 0.039 \\
$\Mgone$ & 181 & 0.006 & \\
$\Mgtwo$ & 181 & 0.007 & \\
Mgb      & 218 & 0.24 & 0.024 \\
$\Fe$    & 217 & 0.22 & 0.034 \\
Fe5406   & 193 & 0.24 & 0.062 \\
Fe5709   &  31 & 0.28 & 0.165 \\
Fe5782   &  31 & 0.27 & 0.213 \\
NaD      &  25 & 0.32 & 0.031 \\ \hline
\end{tabular}

Note -- $\Fe$=(Fe5270+Fe5335)/2.
\end{table}

The line indices were aperture corrected
to a circular aperture with a metric diameter of $1.19 h^{-1}$\, kpc 
(${\rm H}_0 =100 h\,{\rm km\,s^{-1}\,Mpc^{-1}}$), equivalent to
3.4 arcsec at the distance of the Coma cluster.
Further, the indices were corrected for the effect of the velocity
dispersion, and they were calibrated to consistency with the 
Lick/IDS system. The corrections and calibration are described in 
the Appendix, which also contains tables of the final values.

For galaxies in Coma and DC2345-28 we have used literature data 
($\log \sigma$ and $\Mgtwo$) from Davies et al.\ (1987), 
Dressler (1987), Lucey et al.\ (1991) and Guzm\'{a}n et al.\ (1992).
The data from the literature have been aperture corrected and 
transformed to a consistent system, see Paper I.
The typical measurement errors on the literature data are:
$\log \sigma$, 0.025-0.036; and $\Mgtwo$, $\pm 0.010$.

The magnesium indices Mgb, $\Mgone$ and $\Mgtwo$ are strongly
correlated. The relations between the three indices do not have any 
significant intrinsic scatter, see Appendix.
In the following only $\Mgtwo$ is used, since from an observational
point of view no extra information is contained in Mgb and $\Mgone$.
For 37 galaxies without measured $\Mgtwo$ this index was derived
from the measured Mgb, cf.\ Appendix.

\begin{table}
\caption[]{Wavelength definition for $\HbG$
\label{tab-wavelength} }
\begin{tabular}{lrr}
Index    & Central passband  & Continuum passbands \\ \hline
$\HbG$   & 4851.32 - 4871.32 & 4815.00 - 4845.00 \\
         &                   & 4880.00 - 4930.00 \\ \hline
\end{tabular}
\end{table}

\begin{figure}
\epsfxsize=8.5cm
\epsfbox{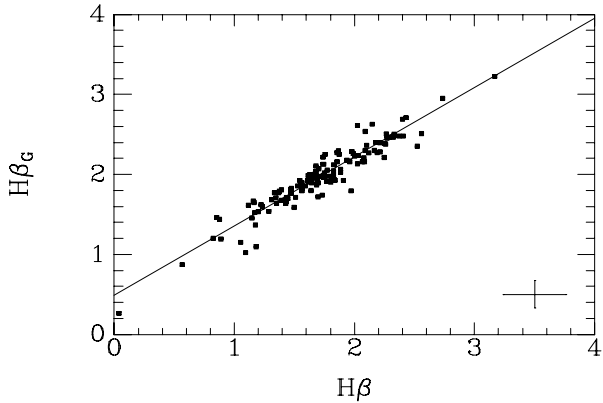}

\caption[ ]{The relation between the Lick/IDS $\Hb$ index and the 
$\HbG$ index.  Typical error bars are given on the figure.
\label{fig-transHb} }
\end{figure}

\subsection{Redefinition of the $\Hb$ index}

The index for $\Hb$ as defined in the Lick/IDS system has
very narrow continuum bands, 20{\AA} and 15\AA. 
This results in a relatively high uncertainty on the derived index. 
We have therefore experimented with a redefinition of the $\Hb$ index.
We have adopted the same wavelength intervals for the $\Hb$ index as 
used by Gonz\'{a}les (1993) for his emission line index, see 
Table \ref{tab-wavelength}.
The index is in the following called $\HbG$.
The central passband of $\HbG$ is narrower than the central 
passband for $\Hb$. This limits the contribution from the iron
line which is within the passband of the Lick/IDS $\Hb$.
Though the central passband for $\HbG$, like for the Lick/IDS
definition, is too narrow to measure the real strength of the
$\Hb$ line in A stars and hotter stars, the wider continuum 
bands give a better relative measure of the line strength.
The main advantage of $\HbG$ is that the median 
uncertainty of log\,$\HbG$ as derived from the S/N ratio of the 
spectra is 0.038, while the median uncertainty of log\,$\Hb$ is 0.064.
Naturally there is a tight correlation between $\Hb$ in the Lick/IDS 
system and the new $\HbG$. For the 129 galaxies with both indices 
measured and both positive we find 
% (delta100 per fit)
\begin{equation}
\arraycolsep=2pt
\begin{array}{ll}
\HbG  = & \hspace*{7.5pt}0.866\,\Hb + 0.485 \\
        & \pm 0.044
\end{array}
\label{eq-Hb}
\end{equation}
with an rms scatter of 0.13, see Figure \ref{fig-transHb}.
Four galaxies have the blue continuum of $\HbG$ outside the wavelength
range of the spectra, while it was possible to measure $\Hb$. 
For these galaxies Eq.\ \ref{eq-Hb} was used to transform $\Hb$ 
to $\HbG$.  In the following analysis $\HbG$ is used in place 
of $\Hb$ and the model predictions of $\Hb$ are transformed to $\HbG$ 
using Eq.\ \ref{eq-Hb}.

\subsection{Emission lines}

Many E and S0 galaxies are known to have emission lines
from ionized gas especially in the central part of the galaxy
(e.g., Davidge 1992; Gonz\'{a}les 1993; Goudfrooij et al.\ 1994).
Galaxies with significant emission from [OIII]$\lambda$5007
and/or $\Hb$ are marked in Table \ref{tab-data1}.
Based on the S/N of the spectra and on comparison with
the data from Gonz\'{a}les (1993) for the galaxies in common,
we judge that emission in [OIII]$\lambda$5007 
with equivalent width larger than $\approx$0.5{\AA} will be detected as
significant emission.  16 galaxies in the cluster sample and 5 galaxies
in the additional sample have significant emission.

The line indices $\HbG$, Fe5015, and to a smaller extent Mgb
can be affected by emission.
When [OIII] emission is present also emission in $\Hb$ is present
and will fill up the stellar $\Hb$ absorption line.
The Fe5015 index has [OIII]$\lambda$5007 in the central passband 
and [OIII]$\lambda$4959 in the blue continuum passband.
Since [OIII]$\lambda$5007 is the stronger of the oxygen lines
the index for Fe5015 will be weakened by the emission.
The Mgb red continuum passband contains [NI]$\lambda$5198,5200.
These lines are significantly weaker than [OIII]$\lambda$5007.
Goudfrooij \& Emsellem (1996) 
find [NI]$\lambda$5198,5200/[OIII]$\lambda$5007$\approx$0.4 for NGC2974.
The effect of the emission is to make Mgb artificially stronger.
We did not attempt to correct any of the line indices for
emission. Instead galaxies with significant emission are omitted
when relations that involve the $\HbG$ index are established.

\section{The empirical point of view}

An important question regarding galaxy evolution is which parameters
determine the mix of stellar populations in E and S0 galaxies.
In this section it is investigated to what extent the observed
stellar populations of E and S0 galaxies are determined by the
depth of the potential well of the galaxies, the mass of the
galaxies and the cluster environment.
The velocity dispersion is used as a measure of the depth of the 
potential well. Further, we study how well the M/L ratio is correlated
with the stellar populations.

In order to characterize the stellar populations we need indices
that will enable us to detect variations in age, metallicity
and abundance ratios.
We use $\Mgtwo$, $\Fe$ and $\HbG$ as the primary indices.
$\HbG$ is age sensitive, but also sensitive to the presence of
blue horizontal branch stars (e.g., Bressan et al.\ 1996).
Tripicco \& Bell (1995) have studied how the Lick/IDS indices
respond to changes in the abundances of various elements.
They find that the $\Mgtwo$ index and the $\Fe$ index are mostly 
sensitive to the magnesium and the iron abundance, respectively.
However, $\Fe$ is as sensitive to changes in the total 
metallicity as to changes in the iron abundance. 
The only index that reacts stronger to changes in the iron abundance 
than to changes in the total metallicity is Fe4383.
This index cannot be measured from our spectra because of their
limited wavelength coverage.
The Fe5406 index, which is also iron sensitive, is not used because
the relative uncertainty on this index is larger than for $\Fe$,
cf.\ Table \ref{tab-index}.
We briefly discuss the C4668 index and the NaD index.

\begin{table}
\caption[]{Model predictions from Vazdekis et al.\ (1996a)
\label{tab-model} }
\begin{tabular}{l@{\,$\approx$\ }l} \hline
$\Mgtwo$   & \hspace*{8pt}0.12\,log\,age + 0.19[M/H] + 0.14 \\
$\log \Fe$ & \hspace*{8pt}0.12\,log\,age + 0.25[M/H] + 0.34 \\
$\log \HbG$ & -- 0.27\,log\,age -- 0.135[M/H] + 0.51 \\ 
$\log M/L_r$ & \hspace*{8pt}0.63\,log\,age + 0.26[M/H] -- 0.16 \\ \hline
\end{tabular}

Note -- [M/H]$\equiv \log Z/Z_{\odot}$ is the total metallicity 
relative to solar.
\end{table}

Models for single stellar populations like those by 
Vazdekis et al.\ (1996a) relate age and metallicity to expected 
values of the line indices and the M/L ratio.
Table \ref{tab-model} lists approximate relations
derived from these authors'  models with a bi-modal IMF with
a Salpeter-like slope $\mu = 1.35$, and ages of 5Gyr or larger.
The models have solar abundance ratios.
However, to a first approximation we assume that the $\Mgtwo$ index
and the $\Fe$ index measure the magnesium abundance and the iron
abundance, respectively.
The relations in Table \ref{tab-model} are used in the following
to quantify the changes in age and/or metallicity required to
produce the ranges and offsets for the various indices.

\begin{figure}
\epsfxsize=8.5cm

\epsfbox{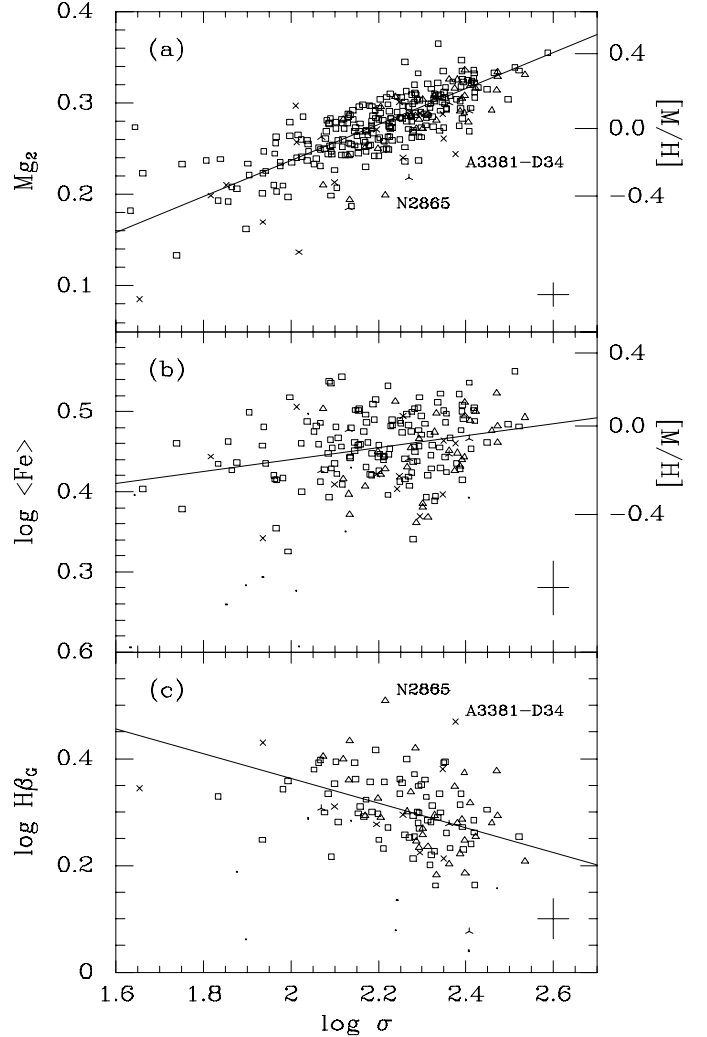}
\caption[ ]{The line indices $\Mgtwo$, $\HbG$ and $\Fe$ versus the
velocity dispersion.
The [M/H] axis on (a) and (b) shows the total metallicity relative
to solar for single stellar population models from 
Vazdekis et al.\ (1996a). The models have a bi-modal IMF with slope
$\mu=1.35$ and an age of 12Gyr.
Open symbols -- galaxies without detected emission lines.
Skeletal symbols -- galaxies with significant emission lines.
Four vertices -- the cluster sample.
Three vertices -- the additional sample.
On (b) and (c) galaxies for which the uncertainty is $\le$0.065 
on $\log \Fe$ and $\log \HbG$, respectively, are shown as open or 
skeletal symbols.  Typical error bars are given on the panels.
Measurements with larger uncertainty are shown as points.
\label{fig-MgHbFesigma} }
\end{figure}

\subsection{The central velocity dispersion and
the stellar populations}

The velocity dispersion is known to correlate strongly with the
$\Mgtwo$ index (e.g., Burstein et al.\ 1988b; Bender et al.\ 1993).
Figure \ref{fig-MgHbFesigma}a shows this relation for the 250 E and S0 
galaxies the cluster sample and for the 40 E and S0 galaxies in the 
additional sample. The relation derived in Paper II is overplotted.
A fit to all the data in the cluster sample gives the same relation,
\begin{equation}
\arraycolsep=2pt
\label{eq-Mgsig}
\begin{array}{ll}
\Mgtwo = & \hspace*{7.5pt}0.196 \log \sigma -0.155   \\
           & \pm 0.009 
\end{array}
\end{equation}
with an rms scatter of 0.025.
The coefficient given here is in agreement with determinations
by Burstein et al.\ and Bender et al.
The relation has an intrinsic scatter of 0.020 in $\Mgtwo$.

The indices $\HbG$ and $\Fe$ are also correlated with the velocity
dispersion.  Both the cluster sample and the additional sample were 
included in the analysis of these correlations.
Because $\Fe$ and $\HbG$ have relatively large measurement errors 
galaxies with uncertainty larger than 0.065 (15\%) in $\log \Fe$ and
$\log \HbG$ are omitted from the analysis.
For $\HbG$ also galaxies affected by emission are omitted.

The correlation between $\Fe$ and $\sigma$ is weak. 
A Spearman rank order test
gives a probability of 0.18\% that the parameters are not correlated.
However, the correlation is driven by galaxies with either low or high
velocity dispersion. If the sample is limited to galaxies with 
$\log \sigma$ in the interval from 2.0 to 2.4, there is no significant
correlation between $\Fe$ and $\log \sigma$.
The galaxies with $\log \sigma$ in this interval have a mean $\log \Fe$
of 0.455 with an rms scatter of 0.040.
This agrees with the result from Fisher et al.\ (1996).
These authors found for a relatively small sample of galaxies
with velocity dispersions in the same interval that $\Fe$ and $\sigma$
were uncorrelated.  For the full range of velocity dispersions we find
\begin{equation}
\arraycolsep=2pt
\label{eq-Fesig}
\begin{array}{ll}
\log \Fe = & \hspace*{7.5pt}0.075 \log \sigma +0.291   \\
           & \pm 0.025 
\end{array}
\end{equation}
with an rms scatter of 0.041. 187 galaxies were included in the fit.
The intrinsic scatter is 0.023. 
The relation is shown in Figure \ref{fig-MgHbFesigma}b.
The rms scatter of $\log \Fe$ for all 187 galaxies is 0.043.
Thus, the $\Fe$-$\sigma$ relation explains very little of the spread
in the $\Fe$ indices.

The [M/H] scales on Figure \ref{fig-MgHbFesigma}a and b
show how $\Fe$ and $\Mgtwo$ depend on the total metallicity, [M/H], 
for models with an age of 12Gyr (Vazdekis et al.\ 1996a).
$\Fe$ and $\Mgtwo$ react the same way to differences in
age, cf.\ Table \ref{tab-model}.
Independent of the actual mean age of the galaxies
the shallow slope of the $\Fe$-$\sigma$ relation compared to the
$\Mgtwo$-$\sigma$ relation therefore shows that
the abundance ratio [Mg/Fe] must change with velocity dispersion.
If the galaxies are coeval then [Mg/Fe] may be
0.4 dex larger for galaxies with a velocity dispersion of
250$\kms$ than for galaxies with a velocity dispersion of 100$\kms$,
all caused by a change in the magnesium abundance.
Since the $\Fe$-$\sigma$ relation may be flat for this interval
of velocity dispersion, any increase in the mean age with
velocity dispersion would have to be balanced with a decrease
in the iron abundance.
So even if the slope of the $\Mgtwo$-$\sigma$ relation was caused
by age differences only, there would be an increase in [Mg/Fe]
of $\approx 0.3$ dex between galaxies with $\sigma = 100\kms$ and
galaxies with $\sigma = 250\kms$, caused by a decrease of the
iron abundance.  These results are similar to the results by Peletier 
(1989) and Worthey et al.\ (1992).
Worthey et al.\ found the [Mg/Fe] ratio to be larger than solar for 
galaxies with strong absorption lines.

The $\HbG$ index and the velocity dispersion show a strong correlation.
A Spearman rank order test gives a probability of $< 0.01\%$ that
$\sigma$ and $\HbG$ are uncorrelated. We find
\begin{equation}
\arraycolsep=2pt
\label{eq-Hbsig}
\begin{array}{ll}
\log \HbG = & \hspace*{0pt}-0.231 \log \sigma +0.825   \\
            & \pm 0.082 
\end{array}
\end{equation}
with an rms scatter of 0.061. 101 galaxies were included in the fit.
The intrinsic scatter is 0.047.  The $\Hb$-$\sigma$ relation derived 
here is in agreement with the relations for E and S0 galaxies found by 
Fisher et al.\ (1995, 1996).

We note at this point, that the E and the S0 galaxies follow the
same relations between the line indices and the velocity dispersion.

In order to study the influence of age variations on the scatter
around the $\Mgtwo$-$\sigma$ relation and the $\Fe$-$\sigma$
relation we test if the galaxies that show emission lines
are offset relative to the rest of the sample.
The 21 galaxies with emission lines have a median offset in $\Mgtwo$
of $-0.017\pm 0.009$ and a median offset in $\log \Fe$ of
$-0.035\pm 0.022$ relative to the relations given in Eq.\ \ref{eq-Mgsig}
and Eq.\ \ref{eq-Fesig}, respectively.
The emission lines are most likely caused by a young stellar population.
The offsets are consistent with the mean age of these galaxies 
being $\approx 0.2$ dex younger than the bulk of the galaxies.

The residuals for the $\Mgtwo$-$\sigma$ relation show a very weak 
correlation with $\log \HbG$ mostly driven by a few galaxies with very 
strong $\HbG$.
The weak dependence on $\HbG$ may be seen in a direct determination
of a relation between $\Mgtwo$, $\log \sigma$ and $\HbG$.
A least squares fit gives a coefficient for $\HbG$ which is 
significant on the 3$\sigma$ level.
If we instead minimize the sum of the absolute residuals in 
$\Mgtwo$ and determine the uncertainties by a boot strap
procedure then the $\HbG$ is marginally significant. We find
\begin{equation}
\arraycolsep=2pt
\label{eq-MgHbsig}
\begin{array}{ll}
\Mgtwo = & \hspace*{7.5pt}0.209 \log \sigma - 0.056 \log \HbG - 0.173   \\
         & \pm 0.014 ~~~~~~\hspace*{2pt} \pm 0.042
\end{array}
\end{equation}
with an rms scatter of 0.019 in $\Mgtwo$. The rms scatter
of the the $\Mgtwo$-$\sigma$ relation is 0.020 for the same sample.

On Figure \ref{fig-MgHbFesigma} the galaxies NGC2865 and A3381-D34 
are labeled as examples of galaxies with weak $\Mgtwo$
and strong $\HbG$ for their velocity dispersion.
The values of $\log \Fe$ for NGC2865 and A3381-D34 are fairly normal,
0.429 and 0.460 respectively.
If the weak $\Mgtwo$ and strong $\HbG$ are due to young stellar
populations in the galaxies, one expects that the $\Fe$ index is also 
weakened, cf.\ Table \ref{tab-model}.  It is not easy to understand 
why the $\Fe$ index has a fairly normal strength.
This may indicate that some important clue is missing in our 
present interpretation of these indices.

The residuals for the $\Mgtwo$-$\sigma$ relation are neither
correlated with $\log \Fe$ nor with the residuals for the 
$\Fe$-$\sigma$ relation.
A fit of $\Mgtwo$ as function of both $\log \sigma$ and $\log \Fe$
gives a non-significant iron term.
Thus, the residuals for the $\Mgtwo$-$\sigma$ relation seem
unrelated to variations in the $\Fe$ index.

\begin{figure*}
\epsfxsize=16.0cm
\epsfbox{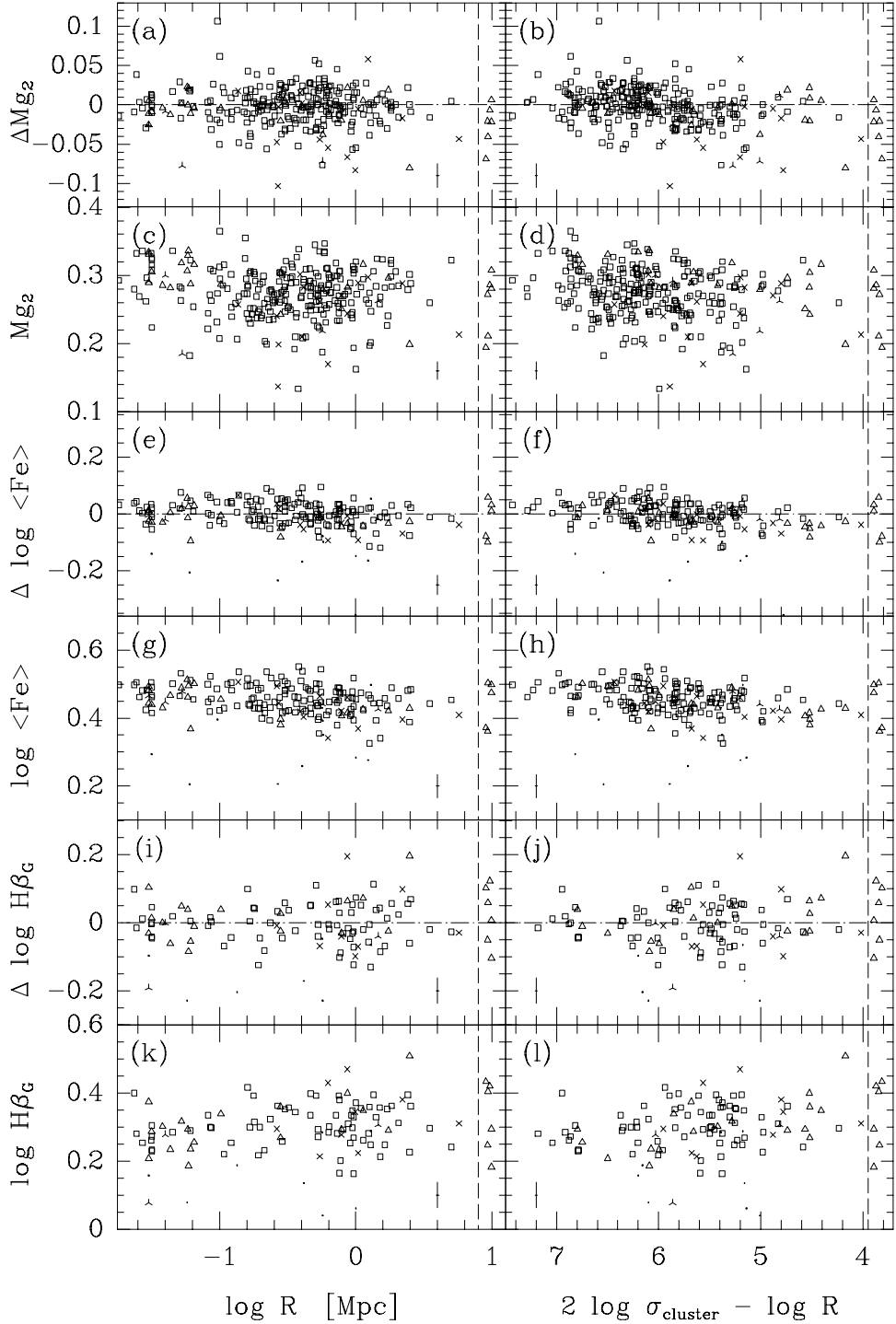}

\caption[ ]{Residuals for the three relations, $\Mgtwo$-$\sigma$,
$\Fe$-$\sigma$ and $\HbG$-$\sigma$, as well as the line indices versus 
cluster center distance, $\log R$, and 
versus $\log \rho _{\rm cluster}=2\log \sigma _{\rm cluster} - \log R$.
$\rho _{\rm cluster}$ is an estimate of the local surface cluster 
density.
The central parts of the clusters and the 
high density environments are to the left on the panels.
Symbols as in Fig.\ \ref{fig-MgHbFesigma}.
Field galaxies in the additional sample are plotted at random
$x$-coordinates right of the dashed lines.
\label{fig-res_Rcl} }
\end{figure*}

\begin{figure*}
\epsfxsize=16.0cm
\epsfbox{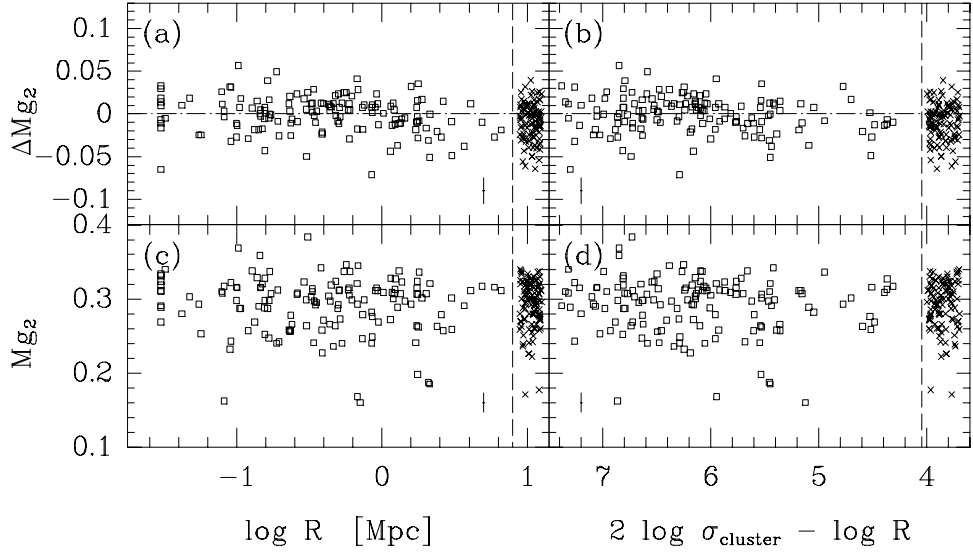}

\caption[ ]{Data published by Faber et al.\ (1989).
The figure shows the residuals for the $\Mgtwo$-$\sigma$ relation and 
the $\Mgtwo$ index versus cluster center distance, $\log R$, and 
versus $\log \rho _{\rm cluster}=2\log \sigma _{\rm cluster} - \log R$.
Boxes -- cluster galaxies; crosses -- field galaxies, plotted at
random $x$-coordinates right of the dashed lines.
\label{fig-7S} }
\end{figure*}

\begin{figure*}
\epsfxsize=17.8cm
\epsfbox{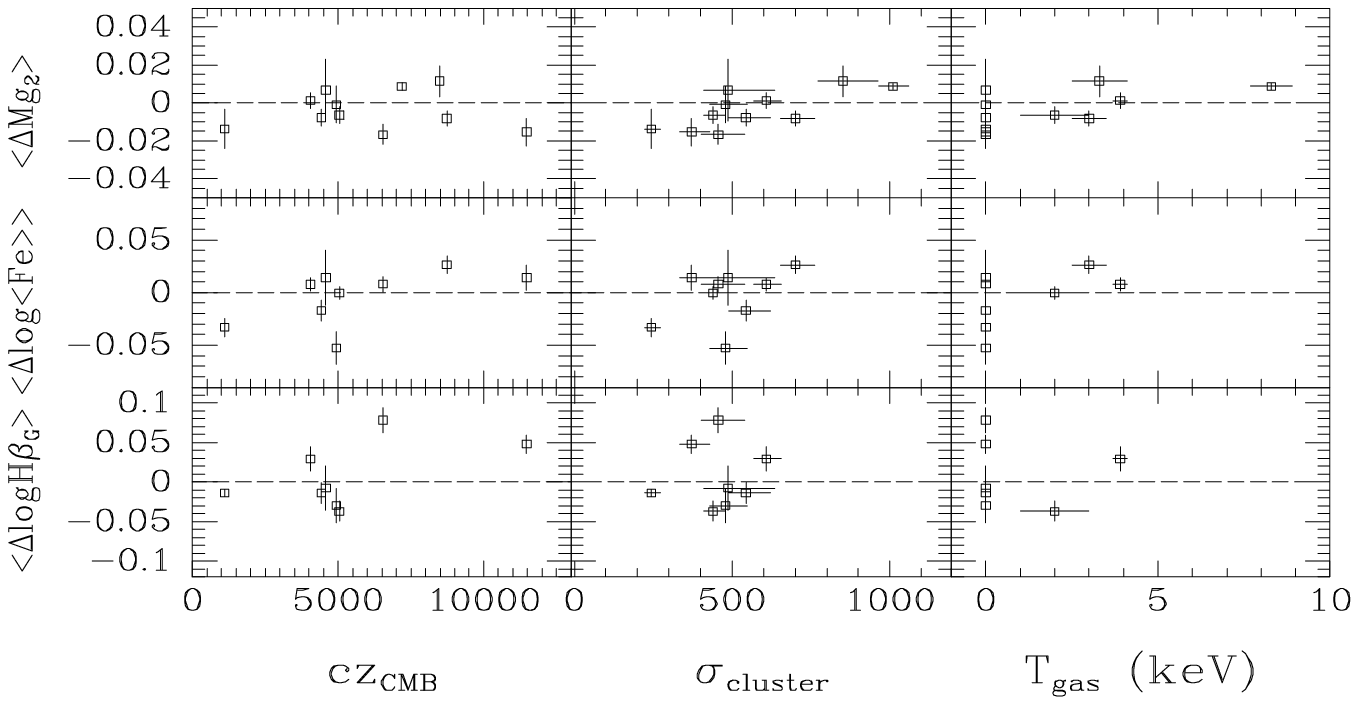}

\caption[ ]{The median zero points for the $\Mgtwo$-$\sigma$ relation, 
the $\Fe$-$\sigma$ relation and
the $\HbG$-$\sigma$ relation versus cluster parameters.
$\sigma _{\rm cluster}$ is the velocity dispersion of the cluster.
$T_{\rm gas}$ is the temperature of the X-ray gas in the cluster.
The values of the cluster parameters are given in Paper II.
\label{fig-zeropoints} }
\end{figure*}

\subsection{Effects of the cluster environment}

In Paper II we found that the residuals for the 
$\Mgtwo$-$\sigma$ relation correlate with the cluster environment,
specifically $\rho _{\rm cluster} =\sigma _{\rm cluster}^2 /R$.
$\sigma _{\rm cluster}$ is the velocity dispersion of the cluster, 
and $R$ is the cluster center distance of the galaxy.
Thus, $\rho _{\rm cluster}$ is an estimate of the projected cluster
surface density. A similar result was earlier found for the Coma cluster
by Guzm\'{a}n et al.\ (1992).

Figure \ref{fig-res_Rcl} shows the residuals for the three relations, 
$\Mgtwo$-$\sigma$, $\Fe$-$\sigma$ and $\HbG$-$\sigma$,
versus cluster center distance and versus $\rho _{\rm cluster}$.
The line indices $\Mgtwo$, $\Fe$ and $\HbG$ are also plotted versus
the cluster environment parameters.
Both the cluster sample and the additional sample are shown.
For 26 of the galaxies in the additional sample the environment
parameters are derived based on cluster velocity dispersions
from Faber et al.\ (1989) and Hickson et al.\ (1992), and redshifts
from Maia et al.\ (1989). The field galaxies in the additional sample 
are separated from the other galaxies on the figure by a dashed line.
These galaxies are not included in the analysis.
None of the results change if all galaxies in 
the additional sample are excluded from the analysis.

There is a weak correlation between $\Mgtwo$ and $\rho _{\rm cluster}$.
We note that the velocity dispersions and $\rho _{\rm cluster}$
are not significantly correlated.
The residuals for the $\Mgtwo$-$\sigma$ relation are
strongly correlated with $\rho _{\rm cluster}$, a Spearman rank order
test gives a probability $< 0.01\%$ that there is no correlation 
(see also Paper II).  This is in agreement with the result found by 
Guzm\'{a}n et al.\ (1992) for the Coma cluster.
For the 276 galaxies with all parameters we find
\begin{equation}
\begin{array}{ll}
\Mgtwo = & \hspace*{7.5pt}0.189\,\log \sigma + 0.009 \log \rho _{\rm cluster}-0.196 \\
         &              \pm 0.012 ~~~~~~~ \hspace*{0.5pt}\pm 0.002
\end{array}
\label{eq-Mgenvironment}
\end{equation}
with an rms scatter of 0.024. The sum of the absolute residuals
in $\Mgtwo$ was minimized. Eq.\ \ref{eq-Mgenvironment} agrees with
our result from Paper II for a smaller sample of galaxies.

The data from Faber et al.\ (1989) for their 10 clusters with most
observed galaxies and reliable cluster velocity dispersions show
the same correlation between the residuals for the $\Mgtwo$-$\sigma$
relation and $\rho _{\rm cluster}$, though the statistical
significance is not as high as for our sample, see Figure \ref{fig-7S}.
If we fit a relation similar to Eq.\ \ref{eq-Mgenvironment} to the
data from Faber et al.\ the coefficient for $\rho _{\rm cluster}$ is
significant on the 2$\sigma$ level.
Our analysis of the Faber et al.\ data included 143 galaxies in
the clusters A194, A2199, Antlia, DC2345-28, Coma, Eridanus, Fornax, 
Perseus, Pisces and Virgo.

Another way of testing the reality of the environment effect of
the $\Mgtwo$-$\sigma$ relation is to compare field galaxies with
cluster galaxies.
The data from Faber et al.\ (1989) provide the largest homogeneous
data base for such a test.
We select as field galaxies all galaxies that according to Faber 
et al.\ are not members of any of the groups listed by these authors. 
Then we test if the residuals for the $\Mgtwo$-$\sigma$ relation
for the 100 field galaxies selected this way and the 143
galaxies that are members of the clusters listed above
are drawn from distributions with the same mean.
A Mann-Whitney rank order test (van der Waerden 1969) gives
a probability of only 0.3\% that this is the case.
The difference in the median zero points for the two samples
of galaxies is 0.009$\pm$0.003. The field galaxies have slightly 
weaker $\Mgtwo$ than the cluster sample.
The field galaxies are included on Figure \ref{fig-7S}, right of 
the dashed lines.  See Burstein, Faber \& Dressler (1990) and 
de Carvalho \& Djorgovski (1992) for other discussions of the
$\Mgtwo$ index for field and cluster galaxies.

The residuals for the $\Fe$-$\sigma$ relation correlate with
both the cluster center distance and $\rho _{\rm cluster}$.
The same is the case for $\Fe$.
A Spearman rank order test gives a probability $<0.01\%$ that
$\Fe$ and $\rho _{\rm cluster}$ are uncorrelated.
In fact the $\Fe$ index correlates stronger with $\rho _{\rm cluster}$
than with any of the local parameters (velocity dispersion, M/L ratio
and mass).  The correlation is not weakened by exclusion of galaxies 
with $\log \sigma$ outside the interval 2.0--2.4.
A relation between $\Fe$ and $\rho _{\rm cluster}$ also has slightly
lower scatter than the $\Fe$-$\sigma$ relation.
If we include both the velocity dispersion and $\rho _{\rm cluster}$
we find the following relation for the 174 galaxies with all parameters
available and the uncertainty on $\log \Fe$ smaller than 0.065,
\begin{equation}
\begin{array}{ll}
\log \Fe = & \hspace*{7.5pt}0.074\,\log \sigma + 0.021 \log \rho _{\rm cluster}+0.170 \\
         &              \pm 0.018 ~~~~~~~ \hspace*{0.5pt}\pm 0.004
\end{array}
\label{eq-Feenvironment}
\end{equation}
The sum of the absolute residuals in $\log \Fe$ was minimized.
The rms scatter of the relation is 0.038.

The residuals for the $\HbG$-$\sigma$ relation as well as $\HbG$
show no significant correlations with the environment 
(Figure \ref{fig-res_Rcl}i-l).  We note, however, that 
galaxies with emission lines favor low density environments.
A relation between $\HbG$, the velocity dispersion and 
$\rho _{\rm cluster}$
have a coefficient for $\log \rho _{\rm cluster}$ of $-0.011\pm 0.009$.

Figure \ref{fig-zeropoints} shows the median zero points for
the relations between line indices and the velocity dispersion for
each cluster versus cluster parameters.
The slightly different values of \mbox{$<\Delta \Mgtwo >$} compared 
to the similar figure in Paper II are due to the fact that
the present work includes more galaxies.
This does not change the result found in Paper II: A Kendall's
$\tau$ correlation test shows that the correlation between
$<\Delta \Mgtwo >$ and the cluster velocity dispersion is 
significant on the 96\% level.
Similar tests were performed for the median zero points for the 
$\Fe$-$\sigma$ relation and the $\HbG$-$\sigma$ relation, but showed
no significant correlations with the cluster parameters.

\subsubsection{Alternative explanations}

Before we study the implications of the detected dependence of
the environment, we discuss if it may be a spurious effect
caused by either the corrections applied to the data or
inconsistencies between data from different sources.
Eqs. \ref{eq-Mgenvironment} and \ref{eq-Feenvironment} show that
the changes in $\Mgtwo$ and $\log \Fe$ between 
$\log \rho _{\rm cluster}$=7.0 and 4.5 are $-0.023$ and $-0.053$, 
respectively.

The velocity dispersion correction of $\Mgtwo$ is very small ($< 0.003$)
and cannot cause the effects.
The adopted aperture correction for $\Mgtwo$ is 
$0.04 \log (r_{\rm ap}/r_{\rm norm})$, cf.\ Appendix.
The correction is largest for galaxies with radial velocities
smaller than 2000$\kms$. Most of these galaxies also happen to be
in low density environments. The mean aperture correction for 
galaxies with $cz_{\rm CMB} < 2000\kms$ and 
$\log \rho _{\rm cluster} < 5.0$ is $-0.024$. If the correct
aperture correction is zero, the adopted correction could in principle
create an offset in $\Mgtwo$ of this size.
However, E and S0 galaxies are know to have radial gradients in
$\Mgtwo$, so a zero aperture correction is highly unlikely.
Further, even if galaxies with $cz_{\rm CMB} < 2000\kms$ are excluded
from the analysis there is a strong correlation between
$\rho _{\rm cluster}$ and the residuals for the $\Mgtwo$-$\sigma$
relation.
The result by Guzm\'{a}n et al.\ (1992) also supports that the
adopted aperture correction did not create a spurious signal,
since these authors' result is based on data for galaxies within
one cluster and therefore does not dependent critically on the
adopted aperture correction.

The velocity dispersion correction of $\Fe$ is significant, 15\% at
$\sigma = 200\kms$ and 32\% at $\sigma = 300\kms$.
A high velocity dispersion makes the raw measurement value of
the index smaller.
The median correction for the galaxies in high density environments
($\log \rho _{\rm cluster} > 6.5$) is 14.5\%, while the median
correction for galaxies in low density environments
($\log \rho _{\rm cluster} < 5.0$) is 10\%.
The maximum difference in $\log \Fe$ between low and high density
environments that this correction can cause is $-0.02$.
The aperture correction used for $\log \Fe$ is 
$0.05 \log (r_{\rm ap}/r_{\rm norm})$, cf.\ Appendix.
The mean correction for galaxies with $cz_{\rm CMB} < 2000\kms$ 
and $\log \rho _{\rm cluster} < 5.0$ is $-0.03$.
Thus, in order to explain the detected change in $\log \Fe$ with
environment as a spurious signal due to these corrections, 
both the velocity dispersion 
correction and the aperture correction for the index need to be zero.
We do not find this very likely to be the case. 
E and S0 galaxies have radial gradients in $\Fe$, so some aperture
correction is necessary. Further, our correction for the velocity
dispersion agrees with similar corrections used by 
Davies et al.\ (1993).
Finally, the galaxies with $cz_{\rm CMB} > 2000\kms$ show just
as strong a correlation between $\Fe$ and $\rho _{\rm cluster}$ as the
full sample.

The adopted aperture corrections for the line indices are based
on average radial gradients of the indices, cf.\ Appendix.
E and S0 galaxies show a significant spread in the 
radial gradients of $\Mgtwo$ and $\log \Fe$ (e.g., Davies et al.\ 1993;
Carollo et al.\ 1993; Fisher et al.\ 1995, 1996).
If the radial gradients depend on the cluster environment then this
may result in a spurious environment dependence for centrally
measured indices aperture corrected with this technique.
Radial gradients of the line indices are not available for the present 
sample, so a direct test cannot be performed.
However, Carollo et al.\ (1993) found for galaxies with masses
smaller than $10^{11}\, {\rm M _{\odot}}$ that the radial gradient
of $\Mgtwo$ was correlated with the galaxy mass, the velocity
dispersion, and possibly with the luminosity and the ellipticity.
If any of these parameters are correlated with the cluster center
distance or with $\rho _{\rm cluster}$ this may also be the case for the
radial gradients of  $\Mgtwo$ (and possibly $\log \Fe$).
For the samples of galaxies used in the present analysis Spearman
rank order tests show no significant correlations between the
cluster environment parameters (cluster center distance and 
$\rho _{\rm cluster}$) and the masses, the absolute luminosities, 
the effective radii in kpc, the velocity dispersions or the 
ellipticities of the galaxies.
Further, there are no significant differences in the median
values of these parameters for the galaxies in high density
environments ($\log \rho _{\rm cluster} > 6.5$) and in low density
environments ($\log \rho _{\rm cluster} < 5.5$).
It should also be noted that the range in the $\Mgtwo$ gradients
for galaxies with masses larger than $10^{11}\, {\rm M _{\odot}}$
is as large as the range for the low mass galaxies 
(cf.\ Carollo et al.\ 1993).
Thus, it does not seem likely that environment dependences for
some of the structural parameters cause spurious environment
dependences for the $\Mgtwo$ and $\Fe$ indices.

The $\Mgtwo$ indices for galaxies in Coma and DC2345-28 are from
the literature. Our data for the clusters HydraI, A539,
A3381 and S639 were mostly taken with the OPTOPUS instrument,
while the rest of our data are from B\&C spectra.
Despite the best efforts there may be inconsistencies between
the literature data, the OPTOPUS data and the B\&C data.
It was therefore tested if the environment dependences can be
detected from subsamples restricted to one or two sources of data.
For the $\Mgtwo$ index we test the literature data alone, our
data alone, and our data divided in OPTOPUS and B\&C data as
outlined above.
The residuals for the $\Mgtwo$-$\sigma$ relation show no dependence
on the environment based on the literature data alone.
This is expected since this subsample consists of the two richest
and most dense clusters, and the range in cluster center
distances is rather limited.
For our data alone a Spearman rank order test gives a 
probability of P=0.6\% that the residuals are uncorrelated with
$\rho _{\rm cluster}$.
For the subsamples of OPTOPUS and B\&C data the probabilities are
1.5\% and 24\%, respectively.  For all subsamples of our data the 
coefficient for $\log \rho _{\rm cluster}$ is consistent with 
Eq.\ \ref{eq-Mgenvironment}.
The comparisons of the $\Mgtwo$ indices from different sources,
see Paper I, did not indicate any large systematic problems.
The results from the subsamples also support that the
environment effect is real.
We will caution though, that we cannot totally rule out that 
part of the detected environment effect for the $\Mgtwo$ index
is spurious and caused by inconsistencies between the
different data sources.

All the determinations of the $\Fe$ index come from our observations.
We divide the sample in two subsamples consisting of the 
OPTOPUS data and the B\&C data, respectively.
The probability that $\Fe$ and $\rho _{\rm cluster}$ are uncorrelated
is 0.8\% for the OPTOPUS data and 0.03\% for the B\&C data.
For both subsamples the coefficient for 
$\log \rho _{\rm cluster}$ is consistent with 
Eq.\ \ref{eq-Feenvironment}.
Because the environment effect for the $\Fe$ index is clearly detectable
in both subsamples, we see no indications that this effect is
caused by inconsistencies in the data.

\subsubsection{Implications of the environment effects}

If the changes in $\Mgtwo$ and $\Fe$ with $\rho _{\rm cluster}$
are interpreted as a change in the abundances only, this implies
that [Mg/H] increases with 0.12 dex and [Fe/H] increases with 0.22 dex,
as the density $\log \rho _{\rm cluster}$ increases from 4.5 to 7 
(cf.\ Table \ref{tab-model}).
This means the abundance ratio [Mg/Fe] must decrease with $\approx 0.1$
dex between $\log \rho _{\rm cluster}=4.5$ and 7.
Because $\Mgtwo$ and $\log \Fe$ are equally sensitive to
age variations it is not possible with the current models
to avoid a decrease in [Mg/Fe] for high density environments
relative to low density environments.
Even if the full change in $\Mgtwo$ is caused by age variations
[Fe/H] must increase with $\approx 0.1$ dex, thus giving the same
decrease in [Mg/Fe] as found above.
We note, however, that it not likely that the change in
$\Mgtwo$ is caused by age variations only, since this would
lead to a too large change of the M/L ratio of the galaxies
as function of the environment (see Paper II). 
Further, the required increase in age of 0.18 dex would give
a change in $\log \HbG$ of $\approx -0.05$.
This is consistent with the data, though only marginally.

The environment dependence we find for the $\Mgtwo$ index and
the $\Fe$ index adds to the growing evidence that the stellar 
populations in E and S0 galaxies are influenced by the
surrounding environment (e.g., Guzm\'{a}n et al.\ 1992;
de Carvalho \& Djorgovski 1992; Rose et al.\ 1994).
The real picture of the environmental effects is most likely
much more complex than a simple gradient with cluster center
distance and/or $\rho _{\rm cluster}$.
The analysis presented here does not take sub-clustering into
account; an effect know to be present and related to recent
star formation in galaxies in the Coma cluster (Caldwell et al.\ 1993).
There may also be differences between clusters of similar richness.

It should emphasized that the galaxy sample used for this analysis is by
no means complete and the results should be tested for a complete sample
of galaxies.  It would be very valuable to get $\Fe$ and $\HbG$ for 
galaxies in Coma and DC2345-28, the two richest clusters in our sample.

\begin{figure}
\epsfxsize=8.5cm
\epsfbox{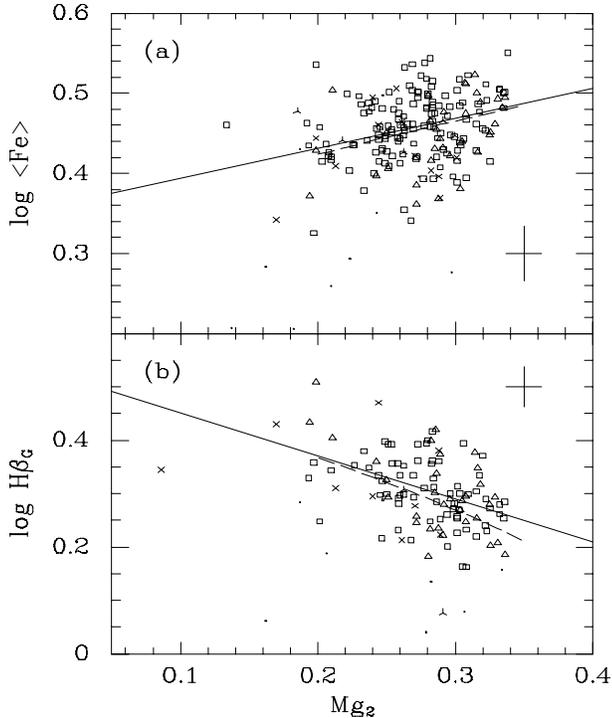}

\caption[ ]{The relations between $\Mgtwo$, $\HbG$ and $\Fe$.
Symbols as in Fig.\ \ref{fig-MgHbFesigma}.
Typical error bars are given on the panels.
Solid lines -- relations derived here.
$\log \Fe = (0.37\pm 0.11)\Mgtwo + 0.36$, 187 galaxies, rms=0.041.
The intrinsic scatter is 0.022.
$\log \HbG = (-0.81\pm 0.20)\Mgtwo + 0.53$, 101 galaxies, rms=0.059.
The intrinsic scatter is 0.044.
Dashed lines -- relations from Burstein et al.\ (1984).
\label{fig-MgHbFe} }
\end{figure}

\subsection{Relations between line indices}

Since all three indices $\Mgtwo$, $\Fe$ and $\HbG$ are correlated
with the velocity dispersion it is also expected that they 
are correlated with each other.
A Spearman rank order test gives a 0.17\% probability that
$\Mgtwo$ and $\Fe$ are uncorrelated, and a probability $<0.01\%$
that $\Mgtwo$ and $\Hb$ are uncorrelated.
Because of the shallow slopes and the large scatter 
of these relations they are derived
by minimization of the sum of the absolute residuals
in $\log \Fe$ and $\log \HbG$, respectively.
The same galaxies were included as for the determination of
Eq.\ \ref{eq-Fesig} and Eq.\ \ref{eq-Hbsig}, respectively.
The two relations are shown in Figure \ref{fig-MgHbFe}.
The derived relations agree within the uncertainties with those for E 
galaxies given by Burstein et al.\ (1984), see Figure \ref{fig-MgHbFe}.

In Sect.\ 4.1 we showed that the correlation between $\Fe$ and the
velocity dispersion
is driven by galaxies that have either low or high velocity dispersion.
This is also the case for the correlation between $\Fe$ and $\Mgtwo$.
The galaxies with $\log \sigma$ in the interval 2.0--2.4 show no
significant correlation between $\Fe$ and $\Mgtwo$.
This emphasizes that the magnesium abundance and the iron
abundance seem to be only losely connected, in agreement with
their different dependence on the velocity dispersion (Sect.\ 4.1).

The residuals for the $\Fe$-$\Mgtwo$ relation are dependent on the
cluster environment in a similar way as the residuals for the
$\Fe$-$\sigma$ relation. A fit that involves both line indices
and $\rho _{\rm cluster}$ gives
\begin{equation}
\begin{array}{ll}
\log \Fe = & \hspace*{7.5pt}0.26\,\Mgtwo + 0.022 \log \rho _{\rm cluster}+0.261 \\
         &              \pm 0.13 ~~~~~ \hspace*{3.5pt}\pm 0.006
\end{array}
\label{eq-FeMgenvironment}
\end{equation}
for the 174 galaxies with reliable cluster parameters.
The relation has an rms of 0.038 in $\log \Fe$.
This relation confirms the environment dependence of [Mg/Fe] discussed
in Sect.\ 4.2.2. 
Eq.\ \ref{eq-FeMgenvironment} shows that for a given value of 
$\Mgtwo$ the $\Fe$ index increases with the cluster density,
and therefore [Mg/Fe] must decrease.

We do not find any environment dependence of the residuals for
the $\HbG$-$\Mgtwo$ relation.
Because of the slope of the relation variations in the metallicity
will not contribute to the scatter. 
Thus, if the environment dependence of $\Mgtwo$ discussed in
Sect.\ 4.2 is caused by changes in the metallicity then
no environment dependence of
the residuals for the $\HbG$-$\Mgtwo$ relation is expected.

$\Fe$ and $\HbG$ are not correlated; 
a Spearman rank order test gives a 68\% probability that
the parameters are uncorrelated.
The residuals for the $\Fe$-$\Mgtwo$ relation are not significantly
correlated 
with $\HbG$, and the residuals for the $\HbG$-$\Mgtwo$ relation 
are not significantly correlated with $\Fe$.
Based on this a combination of $\Mgtwo$ and $\Fe$ is not expected
to give tighter correlation with $\HbG$ than $\Mgtwo$ alone.
For the sample of 100 galaxies with all indices available
and no significant emission we find
\begin{equation}
\arraycolsep=2pt
\label{eq-Hb3}
\begin{array}{ll}
\log \HbG = & -1.02 \Mgtwo +0.45 \log \Fe +0.39   \\
            & \pm 0.21 ~~~~~\hspace*{2.0pt} \pm 0.23
\end{array}
\end{equation}
The sum of the absolute residuals in $\log \HbG$ were minimized.
The coefficient for $\Fe$ is significant on the 2$\sigma$ level, 
The rms scatter is 0.056, thus
the relation does not improve the scatter significantly. 

Fisher et al.\ (1995) fit the quotient Mgb/Fe5270 as function
of $\Hb$ for a smaller sample of E galaxies.
The signs for the dependence on the magnesium and the iron indices
given in Eq.\ \ref{eq-Hb3} are in agreement with the 
result from Fisher et al. 
Eq.\ \ref{eq-Hb3}, however, shows that the $\Hb$ depends
stronger on the $\Mgtwo$ index than on the $\Fe$ index.
Fisher et al.\ were specifically testing if $\Hb$ was determined
by the ratio Mgb/Fe5270 and therefore explicitly assumed that
the dependences 
of the magnesium index and of the iron index had the same strength.

\begin{table*}
\begin{minipage}{14cm}
\caption[]{Relations for the M/L ratio and the mass \label{tab-MLrel} }
\begin{tabular}{l@{\,}c@{\,}lrlllr}
\multicolumn{3}{l}{Relation} & N$_{\rm gal}$ & rms & rms$_{\rm int}$ & min.coord. & \multicolumn{1}{c}{P} \\ \hline
$\log M/L_r$ & = & $(3.13\pm 0.41)\Mgtwo -0.42$ & 207 
   & 0.14 & 0.11 & $\log M/L_r$ & $< 0.01\%$ \\
$\Mgtwo$     & = & $(0.044\pm 0.004)\log {\rm Mass} -0.205$ & 207 
   & 0.028 & 0.024 & $\Mgtwo$ & $< 0.01\%$ \\
$\log M/L_r$ & = & $(1.37\pm 0.50)\log \Fe -0.16$ & 114 
   & 0.17 & 0.14 & $\log M/L_r$ & 0.2\%\\
$\log \Fe$   & = & $(0.042\pm 0.011)\log {\rm Mass} -0.008$ & 114 
   & 0.041 & 0.023 & $\log \Fe$ & 0.01\% \\
$\log M/L_r$ & = & $(-1.66\pm 0.51)\log \HbG +0.96$ & 67 
   & 0.15 & 0.11 & $\log M/L_r$ & 0.01\% \\
$\log \HbG$  & = & $(-0.043\pm 0.014)\log {\rm Mass} + 0.791$ & 67
   & 0.057 & 0.042 & $\log \HbG$ & 1.3\% \\ \hline
\end{tabular}
\end{minipage}
\end{table*}

\begin{figure*}
\epsfxsize=17.8cm
\epsfbox{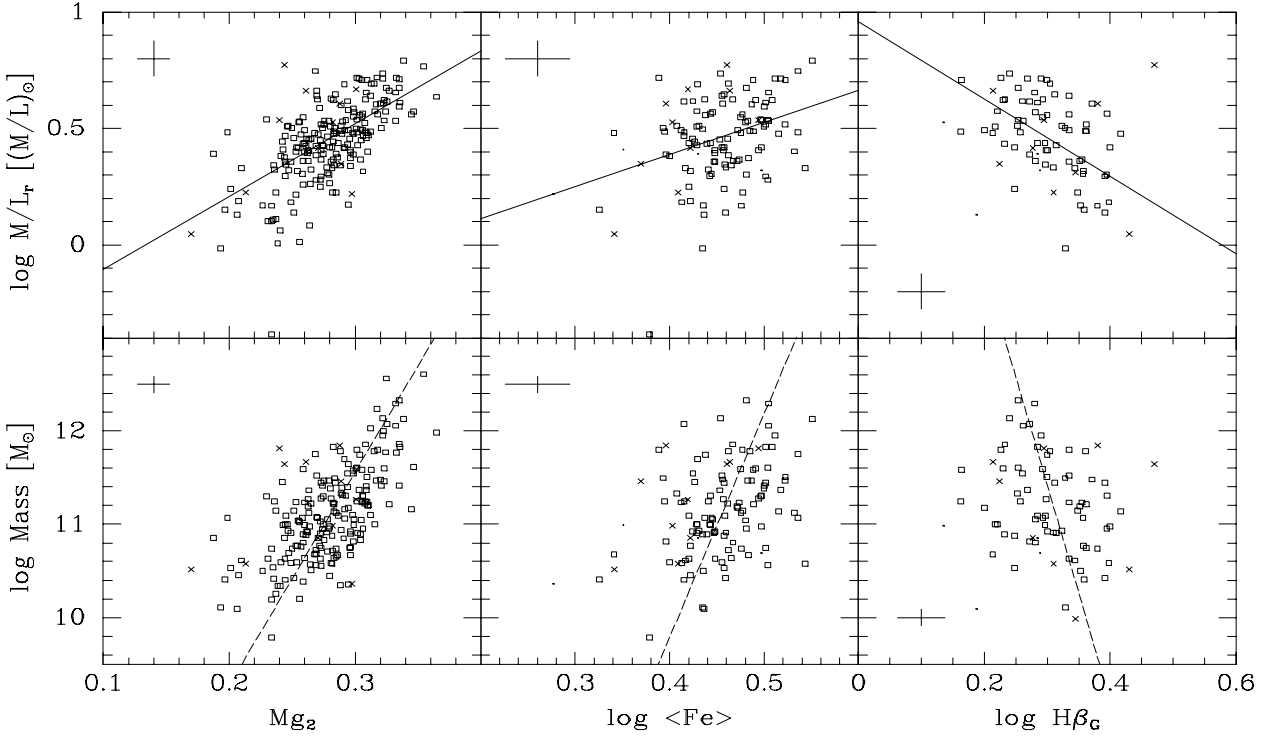}

\caption[ ]{The line indices versus the M/L ratio and the mass of the 
galaxies in the cluster sample. 
Symbols as in Fig.\ \ref{fig-MgHbFesigma}.
Typical error bars are given on the panels.
Solid lines -- relations with residuals minimized in $\log M/L_r$.
Dashed lines -- relations with residuals minimized the line indices.
The relations are listed in Table \ref{tab-MLrel}.
\label{fig-MLMass} }
\end{figure*}

\subsection{The M/L ratio, the mass and the stellar populations}

Next we investigate the correlations between the line indices
and the M/L ratio and the mass of the galaxies.
Only the cluster sample is used for this analysis.
In Paper II we found that for a given mass (or velocity dispersion) the
M/L ratio is determined to 0.1 dex (see also Renzini \& Ciotti 1993).
It must imply that the stellar population is a very strong
function of the mass (or velocity dispersion).
%To investigate this further we ask how well the mass (or the velocity
%dispersion) determines the stellar populations as observed by the
%available line indices.
%And for a given mix of stellar populations, as characterized by 
%the line indices, how well is the M/L ratio determined.

The M/L ratio in Gunn r in solar units is determined as
$\log M/L_r = 2 \log \sigma - \log \Ie - \log \re -0.73$
($\Ho50$), with Mass $=5\sigma ^2 \re /G$ (cf.\ Paper II).
$\re$ is the effective radius in kpc, and $\Ie$ is the mean surface 
brightness within $\re$ in $\rm L_{\odot}/pc^2$.
Because the determinations of the M/L ratios and the masses use
the distance estimates based on the Fundmental Plane (FP)
($\log \re = 1.24\log \sigma - 0.82\log \Ie + \gamma _{\rm cl}$,
see Paper II), it is implicitly assumed that the FP does not
depend on the environment.
This means it will not be possible to detect any environment 
dependence of relations that involve the M/L ratios or the masses,
even though such a dependence is expected for the M/L ratios when
the $\Mgtwo$-$\sigma$ relation and the $\Fe$-$\sigma$ relation 
depend on the environment.

The line indices $\Mgtwo$, $\Fe$ and $\HbG$ are all correlated
with the M/L ratios and with the masses of the galaxies.
The probabilities that the parameters are uncorrelated as derived 
from Spearman rank order tests are given in Table \ref{tab-MLrel}, 
together with the derived relations. The relations are derived by 
minimization of the sum of absolute residuals in either the M/L ratio 
or the line index, as listed in the table.  Figure \ref{fig-MLMass} 
shows the data with the relations overplotted. 
Galaxies with measurement errors larger than 0.065 on $\log \Fe$ and
$\log \HbG$ were omitted from the determinations
of the relations for these line indices.
For relations that involve $\HbG$ galaxies with significant emission
were omitted. 

The correlations of the $\Fe$ index with the M/L ratio and the mass
are, like the correlations of $\Fe$ with the velocity dispersion
and with the $\Mgtwo$ index, driven by galaxies with either low or high 
velocity dispersion.  For the galaxies with $\log \sigma$ in the 
interval 2.0--2.4 the $\Fe$ index is not significantly correlated with 
neither the M/L ratio nor the mass.
Thus, for these galaxies the mass does not determine the $\Fe$ index, 
and the $\Fe$ index does not influence the M/L ratio.
On the other hand, the correlations between the $\Mgtwo$ index
and the M/L ratio and the mass are very tight.
This shows in agreement with the results
from the previous sections that the $\Fe$ index is not governed by
the same quantities that affect the $\Mgtwo$ index.
The $\Mgtwo$ index is nearly fully determined by the velocity 
dispersion (or the mass) of the galaxies, $\Fe$ is not.

For this sample of galaxies the velocity dispersion and the mass 
of a galaxy are equally good determinators of the $\HbG$ index,
since the scatter of the two relations is approximately the same.
The M/L ratio is strongly affected by $\HbG$; as expected if a strong
$\HbG$ is caused by a young stellar population.

\begin{figure}
\epsfxsize=8.5cm
\epsfbox{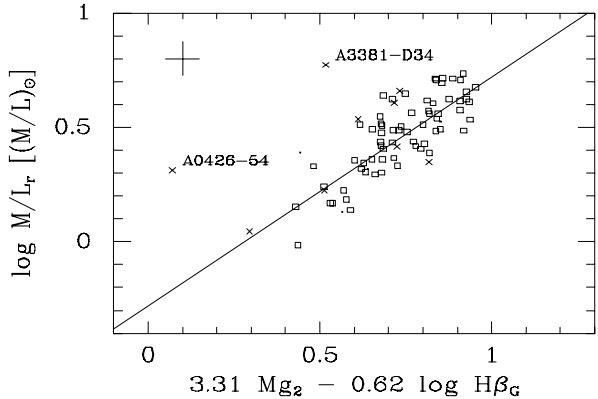}

\caption[ ]{The relation between the M/L ratio and the line indices
$\Mgtwo$ and $\HbG$.
Only galaxies in the cluster sample are shown.
Symbols as in Fig.\ \ref{fig-MgHbFesigma}.
Typical error bars are given on the figure.
Galaxies with emission lines were not included in the determination
of the relation.
\label{fig-MLMgHb} }
\end{figure}

All the relations for the M/L ratio have significant 
intrinsic scatter, cf.\ Table \ref{tab-MLrel}.
The scatter in the M/L ratio at a given line strength is 
0.14--0.17 dex, the corresponding intrinsic scatter is 0.11--0.14 dex.
For the sample of 67 galaxies without significant emission and
for which good determinations of all the parameters are available 
we have tested if a combination of the three line indices will give a 
tighter correlation with the M/L ratio than the relations given
in Table \ref{tab-MLrel}.
A combination of $\Mgtwo$ and $\Fe$ gives a non-significant
iron term and the scatter in the M/L ratio is the same as for
the relation between the M/L ratio and $\Mgtwo$.
A combination of $\Fe$ and $\HbG$ does give significant
coefficients for both terms, but the scatter is not improved
compared to relations that involve only one of the indices.
The best relation is for a combination of $\Mgtwo$
and $\HbG$. We minimize the sum of the absolute residuals
in the M/L ratio and find
\begin{equation}
\begin{array}{ll}
\log M/L_r = & \hspace*{4.5pt}3.31 \Mgtwo - 0.62 \log \HbG - 0.28 \\
             & \hspace*{-3pt} \pm 0.41 ~~~~~\hspace*{2pt} \pm 0.26
\end{array}
\label{eq-MLHbMg}
\end{equation}
with an rms scatter in $\log M/L_r$ of 0.097. 
The relation is shown in Figure \ref{fig-MLMgHb}.
The intrinsic scatter is very small, formally 0.035 dex in $\log M/L_r$.
We stress that the measurement errors may not be determined
accurately enough for this intrinsic scatter to be significant.
For the same sample of galaxies the M/L-$\Mgtwo$ relation has
an rms scatter in $\log M/L_r$ of 0.11 dex and an intrinsic
scatter of 0.07 dex.
Thus, the improvement in the scatter is small and the coefficient for 
$\log \HbG$ is significant only on the 2.5$\sigma$ level.
An interesting property of Eq.\ \ref{eq-MLHbMg} is that variations
in the mean age at a given metallicity will move the data points
nearly parallel to the relation and not contribute significantly to the
scatter. This result is based on the models from 
Vazdekis et al.\ (1996a), see Table \ref{tab-model}.
Variations in the metallicity, [M/H], will contribute to the scatter,
because a change in [M/H] at a given age gives a three times 
larger change in $3.31\Mgtwo - 0.62\log \HbG$ than in the M/L ratio.

\subsection{Other light elements: C and Na}

\begin{table*}
\begin{minipage}{13.3cm}
\caption[]{Relations for C4668 and NaD \label{tab-otherrel} }
\begin{tabular}{l@{\,}c@{\,}lrlllr}
\multicolumn{3}{l}{Relation} & N$_{\rm gal}$ & rms & rms$_{\rm int}$ & min.coord. & \multicolumn{1}{c}{P} \\ \hline
$\log {\rm C4668}$ & = & $(0.63\pm 0.06) \log \sigma - 0.61$ & 57 
  & 0.092 & 0.078 & perpendicular & $<0.01\%$ \\
$\log {\rm NaD}$    & = & $(0.66\pm 0.18) \log \sigma - 0.91$ & 25 
  & 0.053 & 0.036 & perpendicular & $<0.01\%$ \\
$\log {\rm C4668}$ & = & $(2.02\pm 0.40) \Mgtwo + 0.25$ & 57 
  & 0.087 & 0.071 & $\log {\rm C4668}$ & $0.02\%$ \\
$\log {\rm NaD}$    & = & $(2.53\pm 0.81) \Mgtwo - 0.12$ & 25 
  & 0.069 & 0.052 & $\log {\rm NaD}$ & $<0.01\%$ \\ \hline
\end{tabular}
\end{minipage}
\end{table*}

The line index C4668 is a very strong indicator of
variations in the carbon abundance (Tripicco \& Bell 1995).
The line index NaD is mostly sensitive to variations in
the sodium abundance. It is also affected by interstellar 
absorption within the galaxies.

\begin{figure}
\epsfxsize=8.5cm
\epsfbox{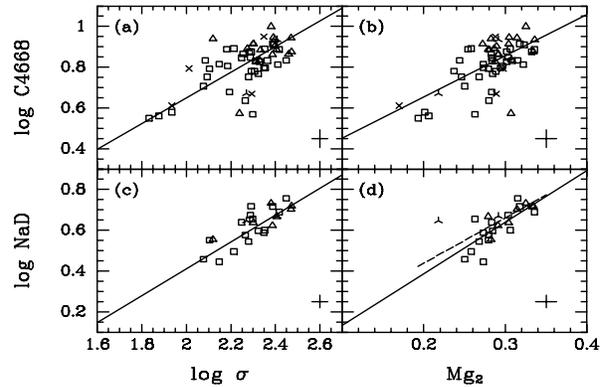}

\caption[ ]{The indices NaD and C4668 versus
the velocity dispersion and the $\Mgtwo$ index.
Symbols as in Fig.\ \ref{fig-MgHbFesigma}.
Typical error bars are given on the figure.
Solid lines -- relations given in Table \ref{tab-otherrel}.
The dashed line on (d) is the relation shown by Burstein et al.\ (1984).
\label{fig-CNaD} }
\end{figure}

Both C4668 and NaD are correlated with the $\Mgtwo$ index and the 
velocity dispersion, while they are not correlated with the $\Fe$ index.
Table \ref{tab-otherrel} lists the derived relations.
The relations are shown in Figure \ref{fig-CNaD}.
The NaD-$\Mgtwo$ relation derived here agrees within the uncertainty 
with the one shown by Burstein et al.\ (1984).

To the extent that the line indices $\Mgtwo$, C4668 and NaD
do measure the abundances of magnesium, carbon and sodium, respectively,
the correlations show that the enrichment of E and S0 galaxies
with these elements are closely connected.
It must await larger samples of galaxies to address if the
intrinsic scatter of the relations given in Table \ref{tab-otherrel}
is significant, and in that case if it is possible to identify
the source of the scatter.

\section{Predictions from stellar population models}

In this section results from static models of single stellar 
populations are used to characterize the average properties of the
stellar populations in the observed galaxies.

We choose, as our basic set of models, the models from 
Vazdekis et al.\ (1996a), which have a bi-modal IMF with a 
Salpeter-like slope of $\mu =1.35$ and solar abundance ratios.
Model ages between 1Gyr and 17Gyr with
metallicities of Z=0.008, 0.02 and 0.04 are available.
The models from Bressan et al.\ (1996) are consistent with
the models from Vazdekis et al.\ that use a Salpeter IMF,
except for the onset of contributions from hot horizontal
branch stars for large ages and high metallicities.
We supplement the models from Vazdekis et al.\ with the high
metallicity (Z=0.1) models from Bressan et al.
%The M/L ratios for these models are from Tantalo et al.\ (1996).
In order to discuss the abundance ratios in more detail
models from Weiss et al.\ (1995) are used. These authors give models
for abundance ratios [Mg/Fe] larger than solar.

None of the conclusions drawn below will change if the
basic set of models had been the Worthey (1994) models, or
the Vazdekis et al.\ models with a Salpeter IMF.
The models from Buzzoni et al.\ (1992, 1994) give weaker 
$\Fe$ and (for [Fe/H]$\ge 0$) stronger $\HbG$ for a given age and 
metallicity, than any of the above mentioned models.
The difference originates partly from differences in the adopted
fitting functions for the line indices.
The models from Buzzoni et al.\ are not used in the following.
Several other stellar population models exist than those 
mentioned here, see Arimoto (1996) and references herein.

\begin{figure*}
\epsfxsize=17.8cm
\epsfbox{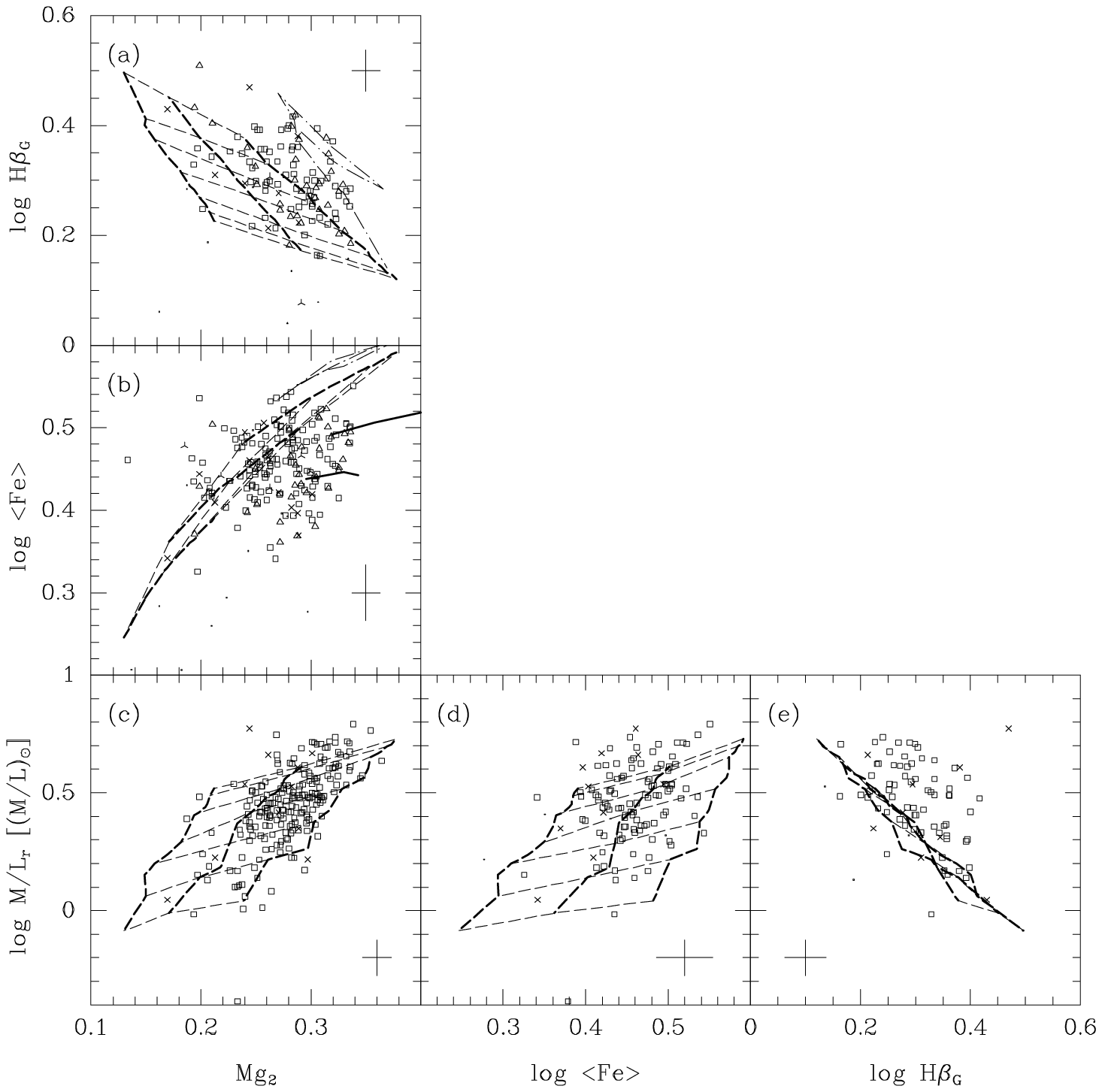}

\caption[ ]{The line indices ($\Mgtwo$, $\HbG$ and $\Fe$)
and the M/L ratio versus each other.
Data symbols as in Figure \ref{fig-MgHbFesigma}.
Typical error bars are given on the panels.
Predictions from static stellar population models are overplotted.
Dashed lines -- Vazdekis et al.\ (1996a), thick lines are constant
metallicity (Z=0.008, 0.02, 0.05), thin lines are constant
age (2, 3, 5, 8, 12, 15, and 17 Gyrs).
Dot-dashed lines -- Bressan et al.\ (1996), Z=0.1, ages from 2Gyr
to 17Gyr.  Solid lines -- Weiss et al.\ (1995) with [Mg/Fe]=0.4 
(Z=0.02 and 0.05, ages from 12Gyr to 18Gyr).
\label{fig-Mod} }
\end{figure*}

Figure \ref{fig-Mod} shows the data with the models overplotted.
Offsets have been applied to the models from Bressan et al.\ (1996) 
such that these authors' model for an age of 15Gyr and Z=0.02
would agree with the similar model from Vazdekis et al.
The models from Weiss et al.\ (1995) for Z=0.02 have been offset in the
same way. The model values for Z=0.05 have been derived by linear
interpolation between the models for Z=0.04 and 0.07 given by 
Weiss et al., and offset to agreement with the Vazdekis et al.\ model
with an age of 15Gyr and Z=0.05.
These offsets ensure that the models from Bressan et al.\ and
Weiss et al.\ can be used together with the Vazdekis et al.\ models
to illustrate the shifts in the observables
expected from very high metallicity or non-solar [Mg/Fe].

\begin{figure*}
\epsfxsize=17.8cm
\epsfbox{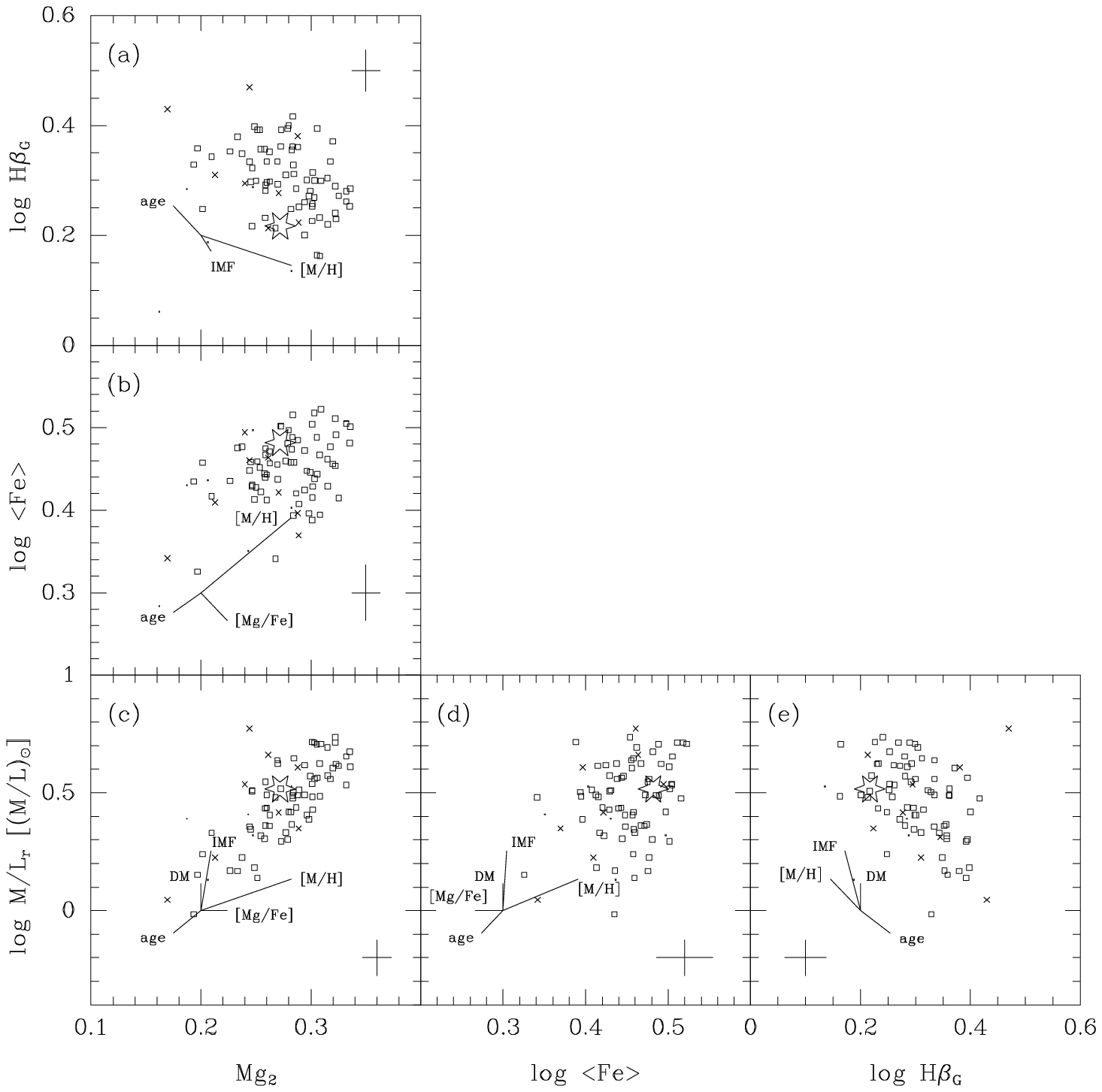}

\caption[ ]{The line indices ($\Mgtwo$, $\HbG$ and $\Fe$)
and the M/L ratio versus each other.
Only galaxies in the cluster sample for which we have all
parameters are plotted, a total of 84 galaxies. 
Data symbols as in Figure \ref{fig-MgHbFesigma}.
Typical error bars are given on the panels.
Large stars -- model values from Vazdekis et al.\ (1996a) for
[M/H]=0, [Mg/Fe]=0, age=12Gyr, and 
slope of the bi-modal IMF $\mu=1.35$.
The lines indicate expected changes in the observables due to
changes in [M/H], [Mg/Fe], age, IMF slope and fraction of
dark matter, see Sect. 5.4.
Note that the indicated change in age is towards younger models.
\label{fig-MLlinetrend} }
\end{figure*}

Figure \ref{fig-MLlinetrend} shows schematically how the observables 
are expected to change due to variations in age, metallicity, abundance
ratio, IMF and fraction of dark matter.
The data on this figure include only the galaxies for which
all the observables are available.

\subsection{Galaxies with strong $\HbG$}

There is a rather large number of galaxies which have strong 
$\Mgtwo$ (larger than 0.3) and a relatively strong $\HbG$, see
Figure \ref{fig-Mod}a.
The models from Vazdekis et al.\ (1996a) indicate that
these galaxies must have very metal rich stellar populations
and also be dominated by very young stars.
In fact, nearly half of the galaxies have stronger $\HbG$ for 
their $\Mgtwo$ than predicted by any model from Vazdekis et al.
The metal rich models from Bressan et al.\ (1996) are able to
predict a strong $\HbG$ for galaxies with strong $\Mgtwo$.
The strong Balmer line comes from the inclusion of hot horizontal
branch stars with high metallicity.
However, according to the models these stars only contribute
significantly to the strength of the Balmer lines for the 
very metal rich model (Z=0.1).

Changes in the IMF slope give only small differences in the
predicted $\HbG$ and $\Mgtwo$ at a given age and metallicity,
see Figure \ref{fig-MLlinetrend}a (Vazdekis et al.\ 1996a). 
Thus, IMF variations are an unlikely source of the large range
of $\HbG$.
Unless blue horizontal branch stars give a significant contribution 
to $\HbG$ in old galaxies, the large range in $\HbG$ cannot easily be 
explained without a similar substantial range in mean ages.

It also appears that a significant fraction of the galaxies 
have mean ages smaller than 5Gyr (see also Worthey et al.\ 1995).
This result is, however, very model dependent.
It should especially be noted that if the galaxies have an 
abundance ratio [Mg/Fe] above solar then age estimates based on 
the $\HbG$-$\Mgtwo$ diagram and models with [Mg/Fe]=0 will be too small.
Determination of mean ages for the individual galaxies is very
uncertain and model dependent, and we will not attempt to do this.

\subsection{Abundance ratios}

The flat $\Fe$-$\Mgtwo$ relation is not predicted by the stellar
population models (Figure \ref{fig-Mod}b). 
This result does not depend on the assumed IMF.
The only known effect that can lead to a nearly constant $\Fe$
while $\Mgtwo$ varies from 0.2 to 0.34 is non-solar abundance ratios,
see Figure \ref{fig-MLlinetrend}b.  Thus, the data and the models 
restate the result by Worthey et al.\ (1992),
that for E and S0 galaxies with strong magnesium lines the
abundance ratio [Mg/Fe] is larger than solar.
The two models from Weiss et al.\ (1995), which we have shown
on Figure \ref{fig-Mod}, have [Mg/Fe]=0.4 and a total metallicity
of Z=0.02 and Z=0.05, respectively.
The abundance ratio [Mg/Fe] is varied by changing the mix of elements 
while keeping the total metallicity constant.
Changing [Mg/Fe] from zero to 0.4 means that the iron abundance is 
decreased with a factor 2.1, while the magnesium abundance is increased
with a factor 1.2. 
Thus, the model with Z=0.02 has [Fe/H]=--0.32 and [Mg/H]=0.08,
and the model with Z=0.05 has [Fe/H]=0.08 and [Mg/H]=0.48.

It is clear from the models that in order to get $\Mgtwo$ larger
than 0.29 the magnesium abundance needs to be above solar.
In the presence of non-solar abundance ratios
a total metallicity above solar is only needed for those
galaxies that also have a relatively strong iron index
($\log \Fe > 0.5$).  We conclude that for solar metallicity
(Z=0.02) the galaxies with $\Mgtwo$ larger than 0.29 must have
[Mg/Fe] between zero and 0.6, a typical value being 0.3.
Galaxies with both strong $\Mgtwo$ (larger than 0.32) and
strong $\log \Fe$ (larger than 0.5) are best fit with models that have 
above solar metallicity (Z=0.05) and [Mg/Fe] around 0.4.
These estimates are in general agreement with results from
Worthey et al.\ (1992) and Weiss et al.\ (1995).
Finally, about half the galaxies with $\Mgtwo$ in the interval 
between 0.2 and 0.29 can be fit with above solar metallicity and solar 
abundance ratios, while the other half must have [Mg/Fe]$> 0$.

\subsection{The M/L ratio, the IMF, and the fraction of dark matter}

Changes in the age or the metallicity of a stellar population naturally
lead to changes in the M/L ratio, see Figure \ref{fig-Mod}c-e.
A change in the slope of the IMF gives a strong change in the M/L ratio.
For the models with ages larger than 5Gyr and a bi-modal IMF an increase
in the slope from $\mu =1.35$ to $\mu =2.35$ results in
a change in log\,M/L of $+0.2$ (see Figure \ref{fig-MLlinetrend}c-e).
The changes for the models with ages 5Gyr or younger are even larger.
For all the models it is, however, expected
that the M/L ratio and $\HbG$ are strongly correlated,
independent of age and metallicity.

In Sect.\ 4.1 we found that the relation between log\,M/L and
$\log \HbG$ has significant intrinsic scatter, 0.1 in log\,M/L.
The relation has nearly the same slope as expected if
all the galaxies have stellar populations with the same IMF.
The scatter cannot easily be explained as due to 
differences in age and/or metallicity.
An increase in the abundance ratio [Mg/Fe] may lead to a decrease in
$\HbG$ (cf.\ Tripicco \& Bell 1995).
If variations in the abundance ratios at a given M/L ratio cause
the scatter in the relation between the M/L ratio and $\HbG$ then
it is expected that the residuals for this relation are correlated
with the residuals for the M/L-$\Mgtwo$ relation and either
anti-correlated or uncorrelated with the residuals for the
M/L-$\Fe$ relation.
However, the residuals for all three relations are correlated with
each other, and the slopes of these correlations are close to one.
Thus, it does not seem likely that variations in the abundance ratio
[Mg/Fe] can cause the scatter in the M/L-$\HbG$ relation.
We are left with two other possibilities. 
Either (1) the IMF of the current stellar population
in the galaxies varies from galaxy to galaxy, in the sense that
galaxies with higher than usual M/L ratio have a steeper IMF,
and galaxies with lower than usual M/L ratio have a shallower IMF.
Or (2) the fraction of dark matter in the galaxies varies.
For both effects it is expected that the residuals for the three
relations between the M/L ratios and the line indices are correlated
with a slope close to one. This is in agreement with the data.
A larger sample of galaxies with $\HbG$ measured with a better
accuracy than for the present data is needed to address the 
significance of the intrinsic scatter in the M/L ratio at a 
given $\HbG$.
It is not clear, however, that it will be possible to distinguish
between the two possible reasons for the scatter outlined above.

\subsection{The general trends}

The diagrams of the M/L ratio versus $\Mgtwo$, $\log \Fe$, and
$\log \HbG$ can together with the indices plotted versus
each other be used as a diagnosis for which effects
govern the evolution of the E and S0 galaxies.
This is illustrated in Figure \ref{fig-MLlinetrend}, which shows
the galaxies in the cluster sample for which all the required 
parameters are available. The large star symbols on the panels
mark the model values from Vazdekis et al.\ (1996a) 
for solar metallicity and abundance ratios,
an age of 12 Gyr, and bi-modal IMF with slope $\mu =1.35$.
Based on the same models the lines on the panels show how changes in 
total metallicity [M/H], abundance ratio [Mg/Fe], mean age, mean IMF,
and fraction of dark matter will affect the parameters.
The length of the lines show changes in [M/H] and [Mg/Fe] from
0.0 to 0.4 at age 12Gyr and IMF slope $\mu =1.35$, in ages from 12Gyr 
to 8Gyr at solar metallicity and IMF slope $\mu =1.35$, and 
in slope of the IMF from $\mu =1.35$ to $\mu =2.35$ at age 12Gyr and 
solar metallicity. 
The change in fraction of dark matter is shown as the change
from adding 30\% mass to the galaxy in the form of dark matter.
The origin of the lines has be arbitrarily shifted from the location 
of the star symbols in order to avoid conflict with the data points.

Ideally we want to derive from the observables 
($\Mgtwo$, $\Fe$, $\HbG$, M/L) the physical parameters
([M/H], [Mg/Fe], age, IMF, fraction of dark matter).
Age,  IMF and abundances should be understood as luminosity weighted
mean parameters for the stellar populations presently observed.
There are five independent physical parameters
and only four observables, excluding any environment parameters.
%Further, the four observables are correlated, and do not give a 
%set of four independent parameters
However, if the fraction of dark matter is not superimposed by
initial conditions during the formation of the galaxy,
but is related to the evolution of the galaxies
in a way that links it to the abundance ratios (specifically [Mg/Fe])
then it may be possible to derive the physical parameters from
the four observables.

It has been suggested that the abundance ratio [Mg/Fe] above solar
is related to the presence of a short period of star formation
early on with an IMF heavily biased towards high mass stars
(Vazdekis et al.\ 1996a; Worthey et al.\ 1992).
Vazdekis et al.\ fit observational data for three nearby E galaxies
with models that have a period of $\approx$1Gyr
where the IMF has a rather flat slope, followed by an
evolution with an IMF with a steeper slope.
No M/L ratios are given for these models. However, if we assume a
conservative average remnant mass of 1M$_{\odot}$ for all
stars with initial mass larger than 2M$_{\odot}$ then
a flat IMF with lower and upper cutoff as used
by Vazdekis et al.\ in the first 1Gyr of a galaxy's evolution
will turn some 10\% of the mass into stellar remnants.
Assuming these remnants at the present age do not contribute
significantly to the luminosity the M/L ratio increases
with a similar amount relative to a model with the same
steep IMF during the whole evolution.
A more stringent test of the idea outlined here requires that
evolutionary models like those by Vazdekis et al.\ (1996a) include
predictions of the evolution of the M/L ratio.

\section{Discussion and conclusions}

We have investigated the stellar populations in E and S0 galaxies
based on spectral line index data for a large sample of
cluster E and S0 galaxies.
The line indices are on the Lick/IDS system, except the index for the 
$\Hb$ line which has been redefined to give a better signal-to-noise.
The indices $\Mgtwo$, $\Fe$ and $\HbG$  are used as the primary indices
to characterize the stellar populations.

Relations were established between the indices and the 
velocity dispersion, the mass, and the M/L ratio of the galaxies.
The relations were used to study which effects determine the current 
stellar populations Also, the influence of the environment has been 
studied.  It is tested whether the M/L ratio depends only on the stellar
populations.  The E and the S0 galaxies follow the same relations.
In the following discussion
the E and S0 galaxies are, therefore, treated as one class of galaxies.
This does not exclude that there are substantial variations in 
the relative disk luminosities (e.g., J{\o}rgensen \& Franx 1994).

We assume that $\Mgtwo$ and
$\Fe$ for a galaxy with a given mean age are related to the 
magnesium abundance and the iron abundance, respectively.
Both indices are also sensitive to the age, and $\Fe$ reacts 
as much to a change in the total metallicity as to a change in 
the iron abundance.
The $\HbG$ is sensitive to the mean age, but also to the relative
contribution from blue horizontal stars and to the metallicity. 

The $\Mgtwo$ index is strongly
correlated with the velocity dispersion and the mass of the galaxies.
The $\Mgtwo$-$\sigma$ relation for E galaxies has long been
well-established (Burstein et al.\ 1988b; Bender et al.\ 1993).
There are galaxies with strong $\HbG$ which have weak
$\Mgtwo$ for their velocity dispersion. Further, $\Mgtwo$
is weakened if the galaxy has emission lines.
Both these effects can be understood if the strong $\HbG$ and/or 
emission are due to a young stellar population.
This possibility for galaxies with weak $\Mgtwo$
was mentioned already by Burstein et al.\ (1988b).
The galaxies with emission lines also have weak $\Fe$ as would
be expected. However, some galaxies with strong $\HbG$ and weak
$\Mgtwo$ have fairly normal $\Fe$. 
It is not clear why this is the case, and it may indicate that
something important is missing in our interpretation of these indices.

The residuals for the $\Mgtwo$-$\sigma$ relation
depend on the environment, specifically 
$\rho _{\rm cluster} = \sigma _{\rm cluster}^2 / R$, which is a measure
of the projected surface density of the cluster. 
$R$ is the cluster center distance.  Galaxies in low
density environments tend to have slightly weaker $\Mgtwo$
for their velocity dispersion than galaxies in high density
environments.
However, the dependence on $\rho _{\rm cluster}$ explains only a 
very small fraction of the intrinsic scatter in 
the $\Mgtwo$-$\sigma$ relation.

The $\Fe$ index is weakly correlated with the velocity dispersion and
the mass of the galaxies. However, for galaxies with velocity dispersion
between 100$\kms$ and 250$\kms$ the $\Fe$ index shows no significant
correlations with the velocity dispersion or the mass.
The $\Fe$ index correlates weakly with the $\Mgtwo$ index, and is
uncorrelated with $\HbG$ index.
The $\Fe$ index is stronger correlated with $\rho _{\rm cluster}$
than with the mass and the velocity dispersion of the galaxy.
Galaxies in low density environments have in general smaller $\Fe$
than galaxies in high density environments.

The $\HbG$ index is correlated with the velocity dispersion, the mass,
and the $\Mgtwo$ index.
The index itself as well as the residuals for the various relations
are not significantly correlated with the cluster environment.

The M/L ratio is strongly correlated with the $\Mgtwo$ index and the 
$\HbG$ index, while the correlation with the $\Fe$ index is very weak.
A relation between the M/L ratio, $\Mgtwo$ and $\HbG$ has a very
low intrinsic scatter. 

The data were compared to single stellar population models
from Vazdekis et al.\ (1996a).
There are three main points from this comparison.
(1) Galaxies with strong $\HbG$ and strong $\Mgtwo$ require
either very metal rich and young stellar populations,
or a significant contribution to $\HbG$ from blue horizontal stars.
(2) The flat $\Fe$-$\Mgtwo$ relation restates the result by
Worthey et al.\ (1992) that many of the strong lined E and S0 galaxies
must have abundance ratios [Mg/Fe] larger than solar.
(3) The intrinsic scatter in the $\HbG$-M/L relation is not 
predicted by the models unless there are variations in
the IMF and/or the fraction of dark matter in the galaxies.

The two first points may be resolved and quantified when 
better stellar population models become available.
The third point, however, cannot easily be resolved.
The difficulty is how to distinguish between IMF differences and
variations in the fraction of dark matter.
If the fraction of dark matter is mostly determined by the formation
process of the galaxy, it may not be possible to break this degeneracy.
If on the other hand the fraction of dark matter is a natural
consequence of the evolution of the galaxy it may be possible
to identify the source of the scatter.

Based on the results summarized above it seems clear that different 
processes must affect the magnesium index and the iron index, and 
maybe also the abundances of these two elements.
We will here discuss this in a little more detail.

Variations in the mean age of the stellar populations are expected
to change $\Mgtwo$ and $\log \Fe$ with the same amounts.
The shallow slope of the $\Fe$-$\sigma$ relation compared to the
$\Mgtwo$-$\sigma$ relation, therefore, shows that the abundance
ratio [Mg/Fe] changes with velocity dispersion.
We estimate that the [Mg/Fe] for galaxies with velocity dispersions
of 250$\kms$ is 0.3 to 0.4 dex larger than for galaxies with
velocity dispersions of 100$\kms$.
If the galaxies are coeval the difference is due to an increase
in the magnesium abundance. Otherwise it is (partly) due to a 
decrease in the iron abundance.
The $\HbG$ index may offer a possibility to distinguish between the two
possibilities, if this index can be used to trace age variations. 
However, most of the current models cannot reproduce the strong $\HbG$ 
seen in some galaxies with strong $\Mgtwo$.
It is also not known how $\HbG$ reacts to changes in [Mg/Fe],
though results from Tripicco \& Bell (1995) indicate that most of the
change in $\HbG$ due to changes in the metallicity is caused
by the magnesium abundance.  Thus, better models may be needed 
before this technique will be feasible.

There are several studies in the literature that indicate that
the stellar populations of E and S0 galaxies are influenced
by the environment (Guzm\'{a}n et al.\ 1992; de Carvalho \& Djorgovski
1992; Rose et al.\ 1994).
The dependence of the $\Fe$ index and the $\Mgtwo$ index on the 
environment we find from our analysis implies 
that [Mg/Fe] is approximately 0.1 dex lower in high density
environments ($\log \rho _{\rm cluster}$=7) compared to low density
environments ($\log \rho _{\rm cluster}$=4.5).
Independent of possible age differences, the decrease is caused
by the iron abundance increasing faster with the cluster density,
than the (possible) increase in the magnesium abundance with cluster 
density.

We note, that literature data show that the $\Mgtwo$ index and
the $\Fe$ index both change within the galaxies.
The average radial gradients of the two indices are consistent with 
a rather small [Mg/Fe] variation within each galaxy 
(Worthey et al.\ 1992; Davies et al.\ 1993; Fisher et al.\ 1995, 1996).

The detection of variations in [Mg/Fe] leaves us with the challenge
of explaining the cause of the variations among the galaxies
and the nearly constant [Mg/Fe] within each galaxy.
Three possible explanations for above solar [Mg/Fe] 
have been mentioned and to some extent
explored by Worthey et al.\ (1992) and Matteucci (1994). 
Different time scales for star formation,
different IMFs, and selective galactic wind loss of metals.
These authors have not addressed the possible effect of the 
cluster environment.
The environment dependence of [Mg/Fe] found in this paper is about
a third of the size of the change related to the velocity dispersion
of the galaxies. 
Thus, it may be important for our understanding of the abundance
ratios that environment effects are taken into account.

\vspace{0.5cm}
Acknowledgements:
This paper benefited from discussions with M.\ Franx.
S.\ M.\ Faber is thanked for making unpublished data available
from the Lick project on spectral line indices for elliptical
galaxies.  G.\ Worthey and R.\ Peletier are thanked for supplying their
stellar population models in computer readable format.
The Danish Board for Astronomical Research and the European Southern 
Observatory are acknowledged for assigning observing time
for this project and for financial support.
This research was supported through Hubble Fellowship grant
number HF-01073.01.94A from the Space Telescope Science Institute,
which is operated by the Association of Universities for Research
in Astronomy, Inc., under NASA contract NAS5-26555.

\appendix

\begin{table*}
\epsfxsize=17cm
\epsfbox{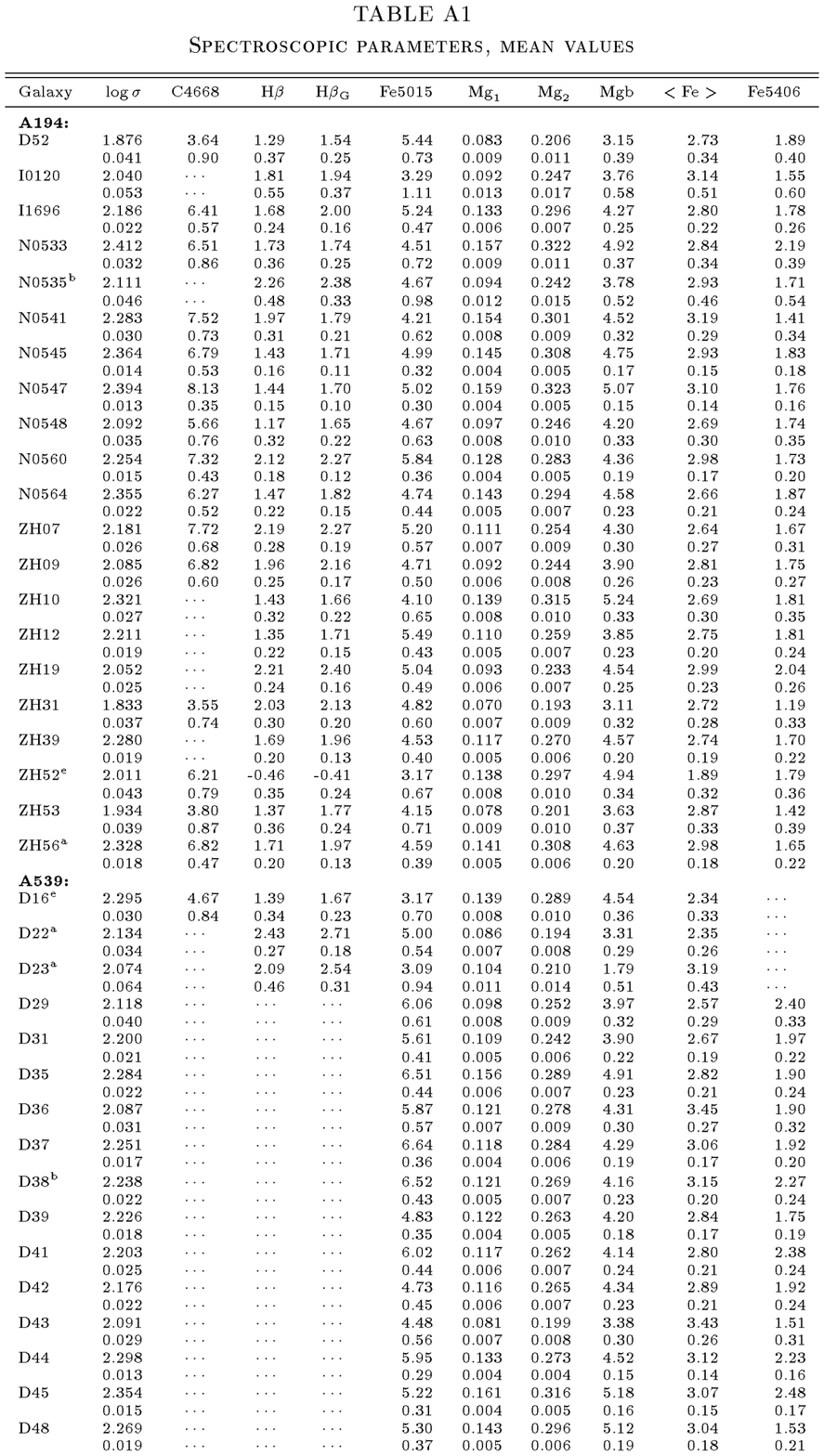}
\end{table*}
\begin{table*}
\epsfxsize=17cm
\epsfbox{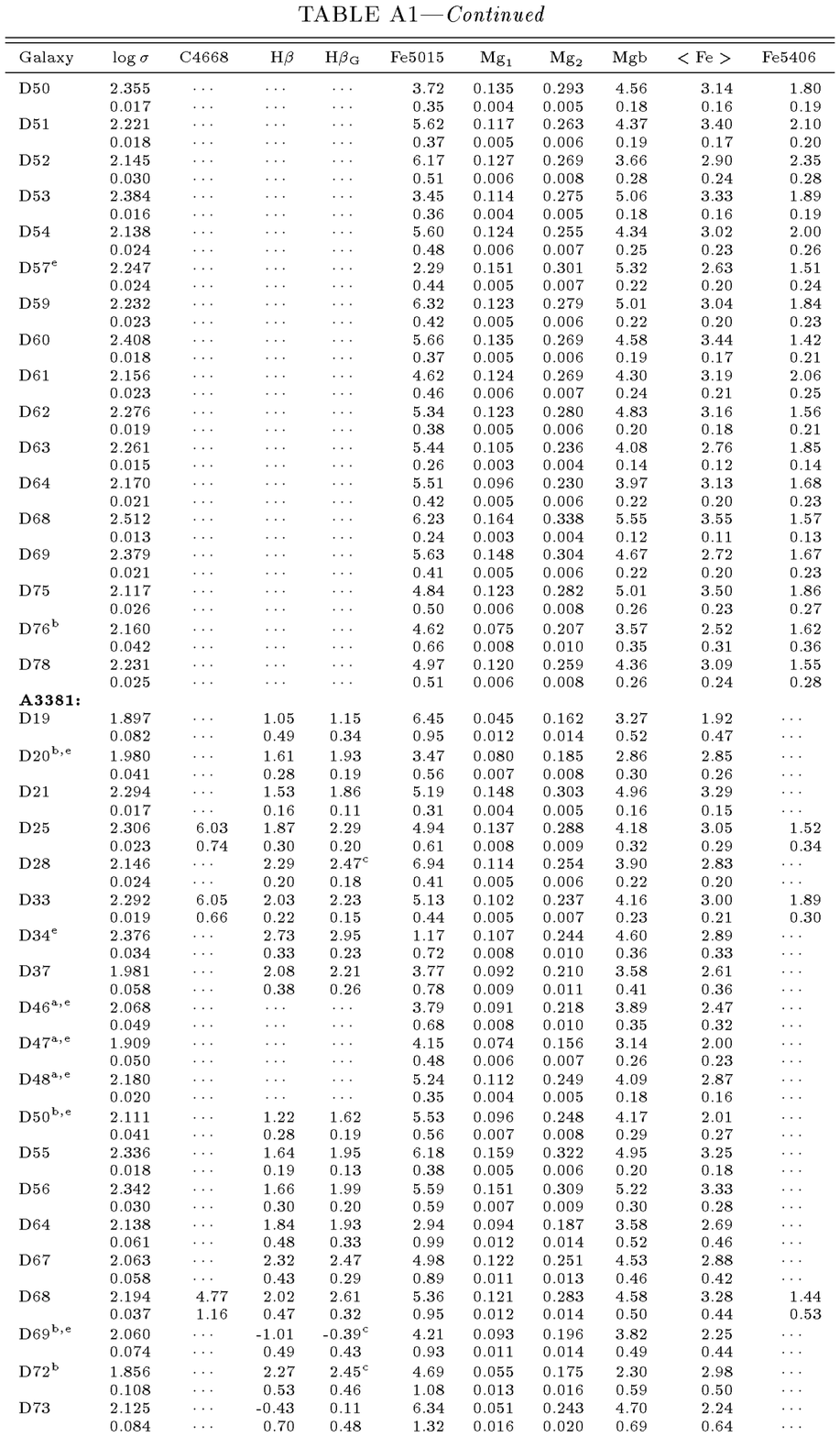}
\end{table*}
\begin{table*}
\epsfxsize=17cm
\epsfbox{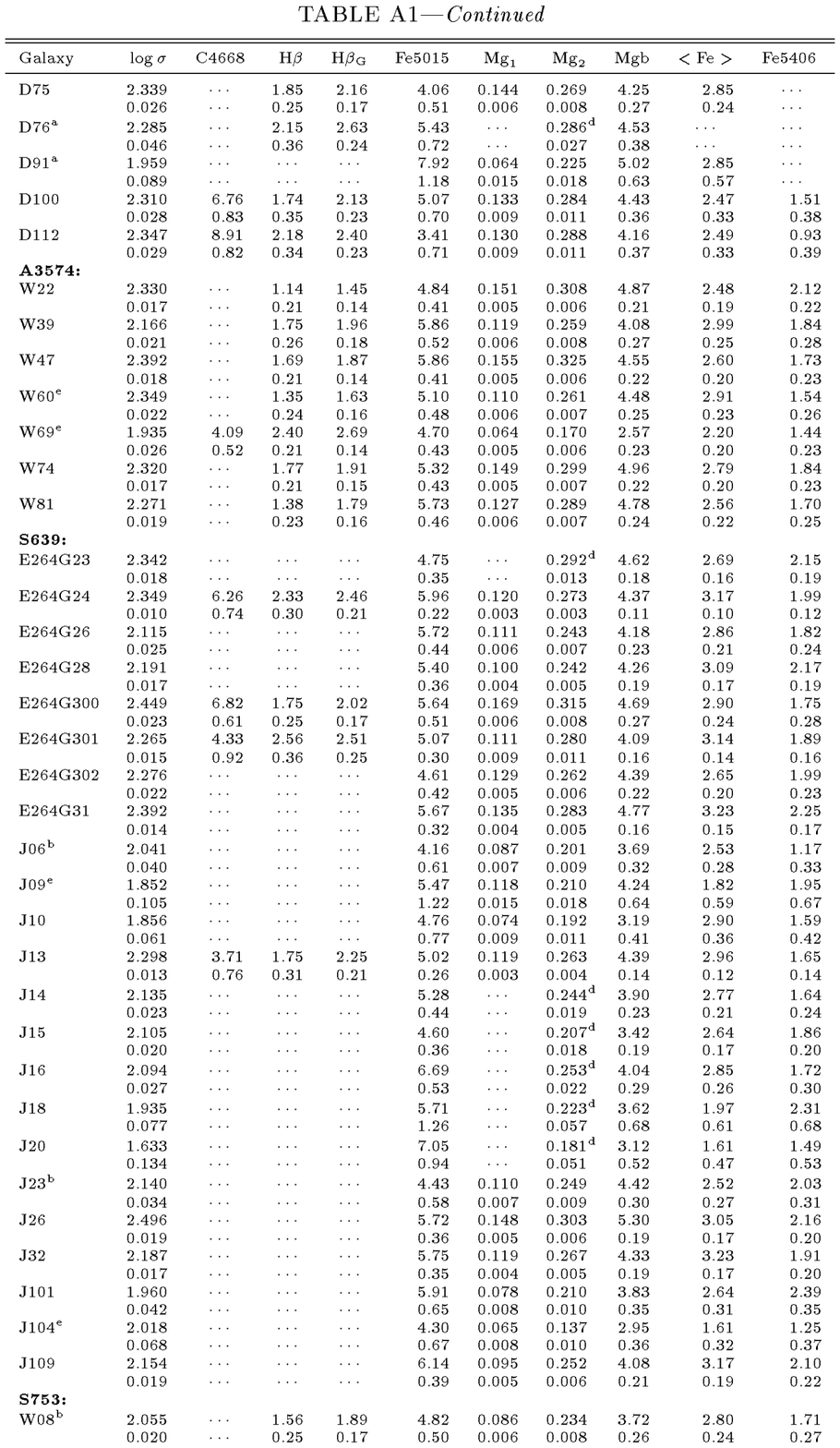}
\end{table*}
\begin{table*}
\epsfxsize=17cm
\epsfbox{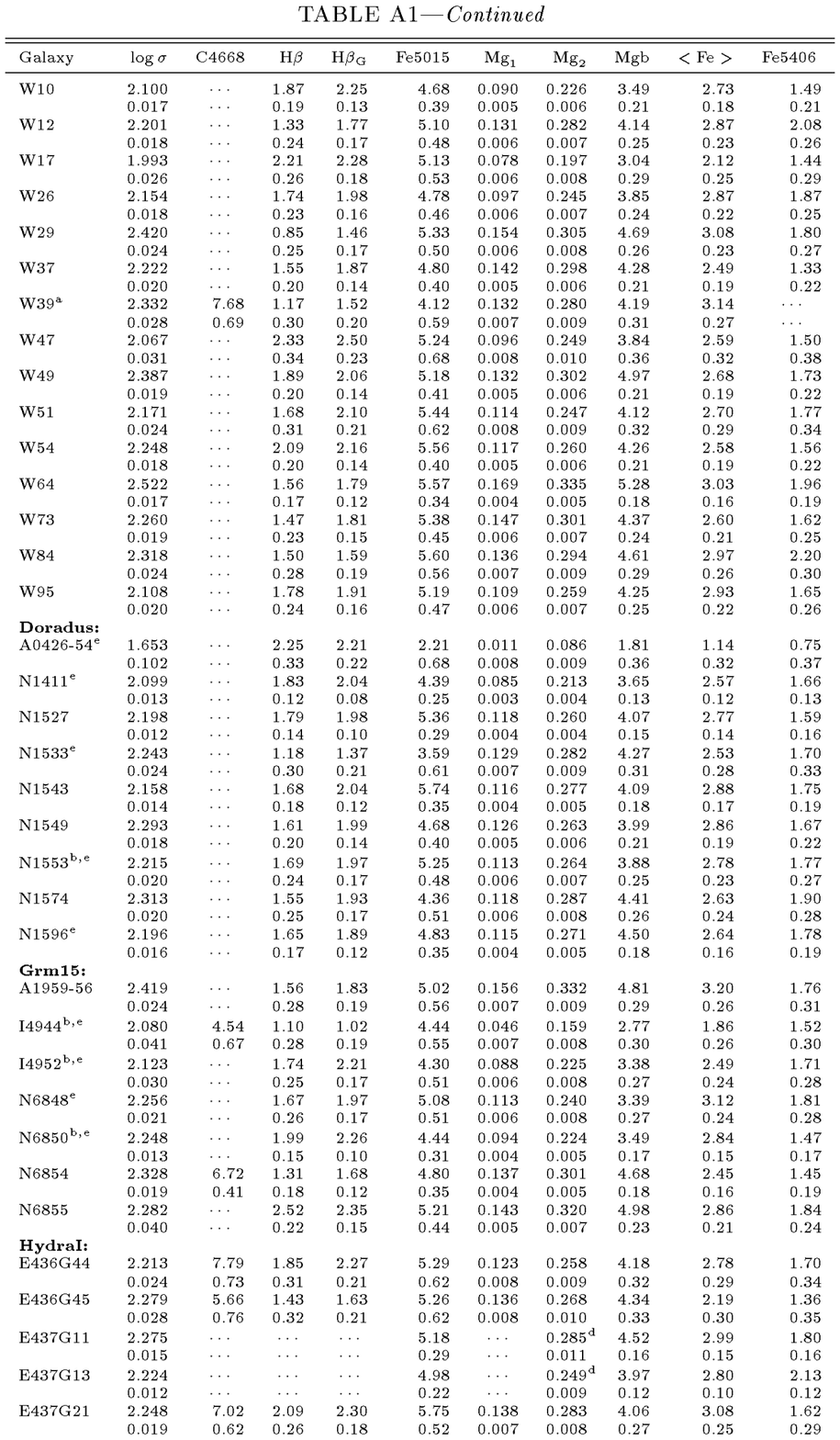}
\end{table*}
\begin{table*}
\epsfxsize=17cm
\epsfbox{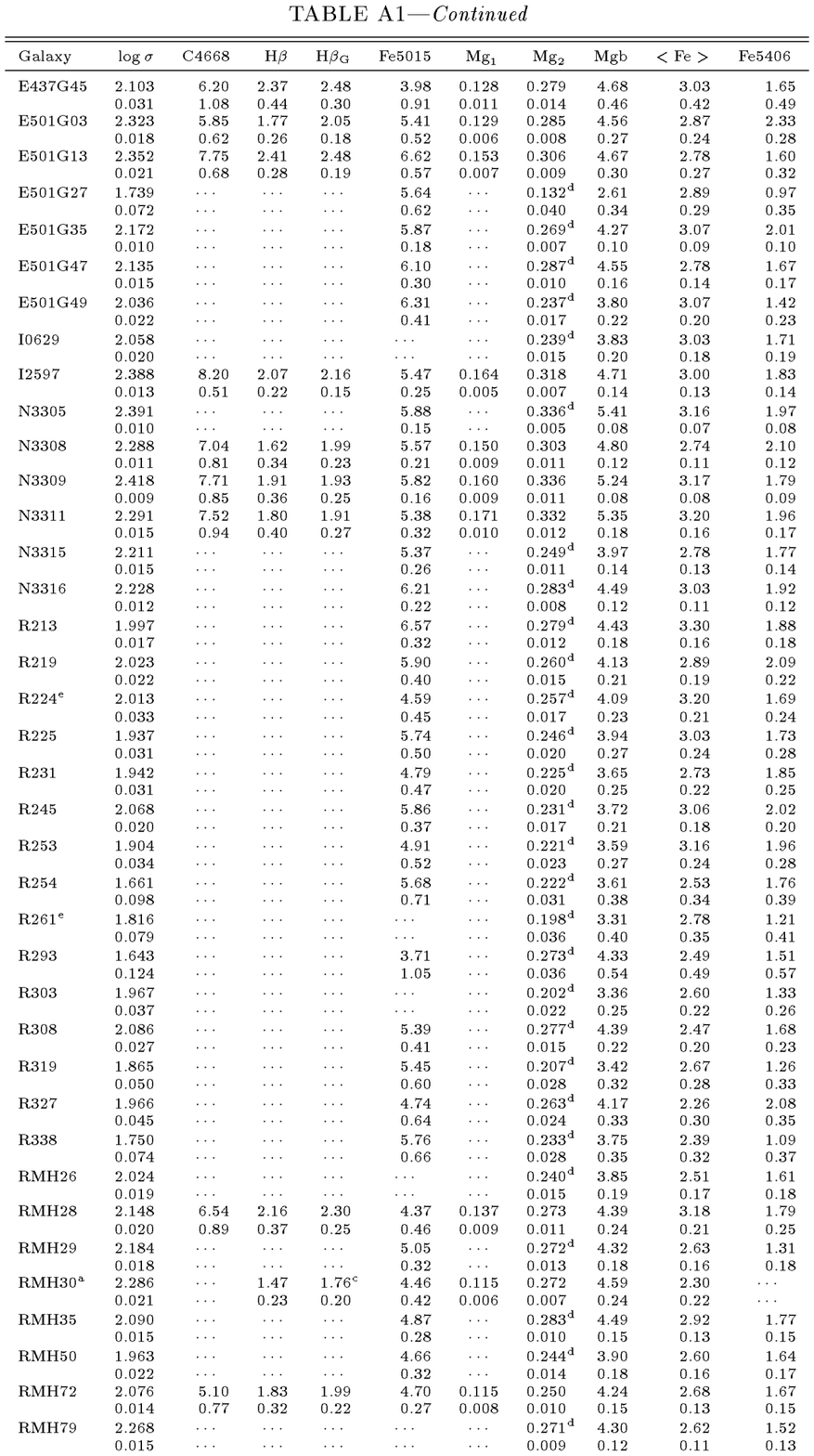}
\end{table*}
\begin{table*}
\epsfxsize=17cm
\epsfbox{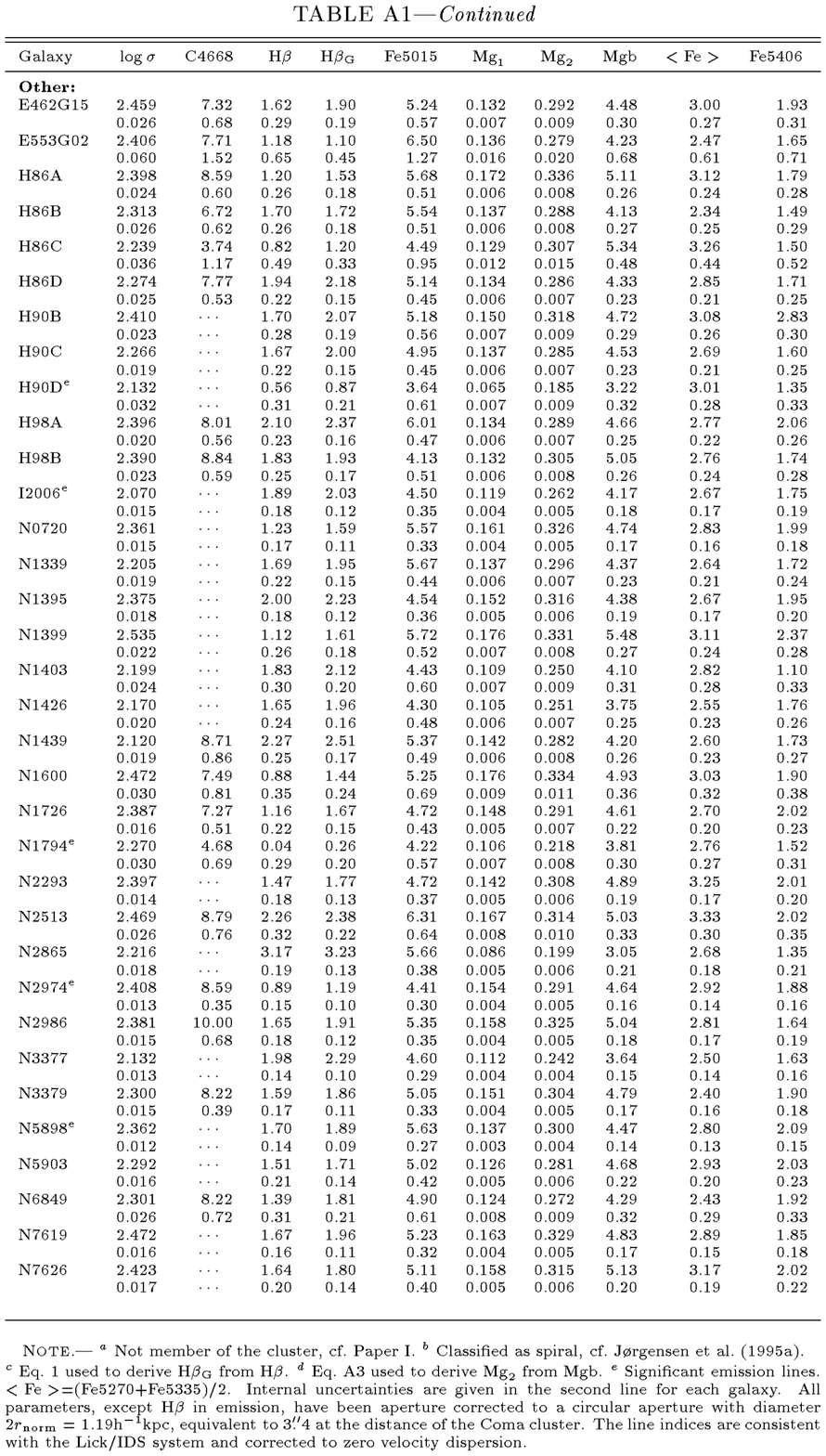}
\end{table*}

\section{Spectroscopic data}

\refstepcounter{table}
\label{tab-data1}
\refstepcounter{table}
\label{tab-data2}

The spectra used in this paper were obtained during three observing
runs at the ESO 1.5m telescope with the B\&C spectrograph and
one observing run at the ESO 3.6m telescope with the OPTOPUS 
instrument.  Information about the instrumentation and the basic 
reductions can be found in Paper I.

\subsection{Determination of line indices}

The line indices were derived from the flux calibrated spectra
convolved to the Lick/IDS instrumental dispersion ($\sigma = 200\kms$,
Gonz\'{a}les 1993).
The B\&C spectra were averaged within the apertures given
in Paper I (B\&C run 1 and run 2 $2\farcs 5\times 8\farcs 16$; 
B\&C run 3 $2\farcs 5\times 6\farcs 8$).
The wavelength intervals for the indices are given by 
Worthey et al.\ (1994) and in Table 2.
All indices except $\Mgone$ and $\Mgtwo$ were derived as
the equivalent width. $\Mgone$ and $\Mgtwo$ are given as 
a magnitude (see also Worthey et al.).

The average values corrected for the aperture size and for
the effect of the velocity dispersion are listed in 
Tables \ref{tab-data1} and \ref{tab-data2}.
For convenience also the velocity dispersion and the $\Mgtwo$
index as published in Paper I are included in Table \ref{tab-data1}.

\subsection{Aperture corrections}

Because E and S0 galaxies have radial gradients in most line indices
the derived 'central' values depend on the distances of the
galaxies and the size of the aperture.
The indices have to be corrected to a standard size aperture.
A circular aperture with diameter
$1.19 h^{-1}$\, kpc was adopted as the standard size. This is equivalent
to 3.4 arcsec at the distance of the Coma cluster.

In general the radial gradients are well described as
logarithmic gradients $\Delta \log {\rm (index)}/\Delta \log r$.
For $\Mgone$ and $\Mgtwo$
$\Delta {\rm Mg}_i /\Delta \log r$ gives a better description, 
cf.\ Fisher et al.\ (1995, 1996).
The radial gradient in the velocity dispersion can also be 
described as a logarithmic gradient (e.g.\ Franx, 
Illingworth \& Heckman 1989).
We therefore assume that the aperture corrections for the 
line indices can be written in the same form as for the
velocity dispersion.  We use
\begin{equation}
\log {\rm (index)_{norm}} = \log {\rm (index)_{ap}} + \alpha \log \frac{r_{\rm ap}}{r _{\rm norm}}
\end{equation}
except for $\Mgone$ and $\Mgtwo$, where we use 
\begin{equation}
{\rm (index)_{norm}} = {\rm (index)_{ap}} + \alpha \log \frac{r_{\rm ap}}{r _{\rm norm}}
\end{equation}
cf.\ Paper I.
$\alpha (\Mgtwo ) = \alpha (\Mgone ) = 0.04$ was used (Paper I).

Gonz\'{a}les (1993) has derived line indices within apertures
with radii $\re /8$ and $\re /2$ for 41 galaxies. 
$\re$ is the effective radius of the galaxy.
From the average values we find that 
$\alpha {\rm (Mgb)} \cong \alpha {\rm (\Fe )} \cong 0.05$ 
and $\alpha {\rm (\Hb )}\cong-0.005$.
We also adopt $\alpha = 0.05$ for all other Fe indices.
Fisher et al.\ (1995, 1996) find the gradients for C4668
(called Fe4668 by these authors) to
be somewhat stronger than for Mgb, Fe5270 and Fe5335.
Mean of all their determinations gives
$\Delta \log {\rm (index)}/\Delta \log r= - 0.11$ for C4668,
while the values for Mgb, Fe5270 and Fe5335 are
$-0.063$, $-0.054$ and $-0.051$, respectively.
We therefore adopt $\alpha {\rm (C4668)} = 0.08$.

Vazdekis et al. (1996) have measured radial gradients of line indices
for 3 galaxies. Their study includes NaD.
They give gradients as $\Delta {(\rm index)} /\Delta \log r$,
and find the gradient for NaD to be significantly stronger
than for Mgb, Fe5270 and Fe5335. Mean for the three galaxies gives
$-1.53$ for NaD. 
The mean gradients for Mgb, Fe5270 and Fe5335 are $-0.88$, $-0.55$
and $-0.50$, respectively.
The strength of NaD is in general comparable to Mgb, while the iron
lines are weaker.  We adopt $\alpha {\rm (NaD)} = 0.09$.

\subsection{Correction for the velocity dispersion}

All the derived line indices, except $\Hb$ in emission, were corrected 
for the effect of the velocity dispersion. 
The corrections were established from K-giant spectra.
The star spectra were convolved to the Lick/IDS instrumental
resolution, and then convolved with Gaussians with $\sigma$
from 50 to 350 $\kms$.  For $\Mgone$ the differences
between the index from the unconvolved spectrum and the 
convolved spectrum as function of the velocity dispersion
was fitted with a low-order polynomial, as done for $\Mgtwo$ in Paper I.
For the other indices the
quotient between the index from the unconvolved spectrum and the 
convolved spectrum was used.
At 200 $\kms$ the corrections are 0.002 for $\Mgtwo$,
2\% for $\HbG$ and 15\% for $\Fe$.
The signs and sizes of the corrections agree with the corrections 
used by Davies et al.\ (1993) for the indices in common.

\subsection{Internal comparison}

\begin{figure*}
\epsfxsize=17.8cm
\epsfbox{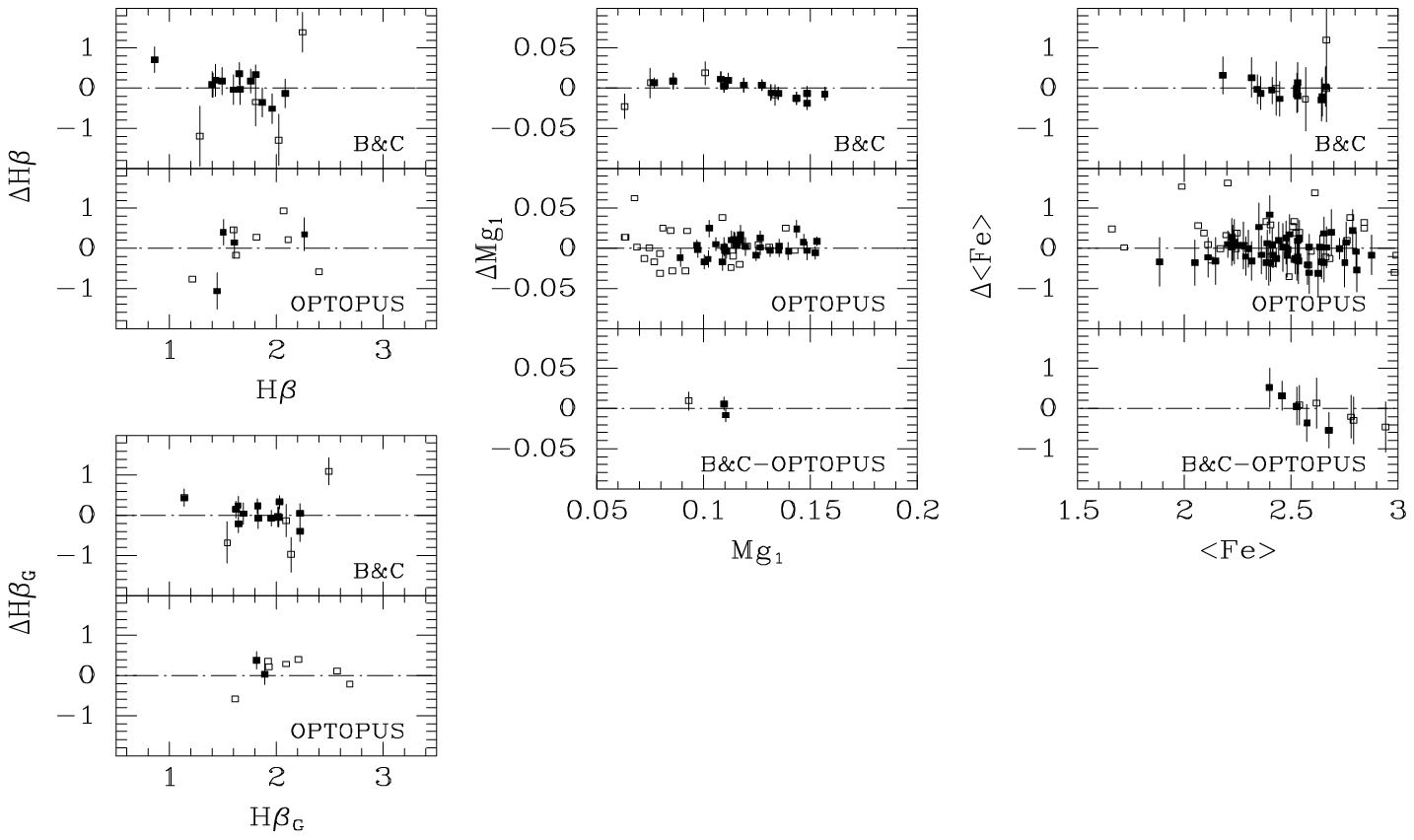}

\epsfxsize=17.8cm
\epsfbox{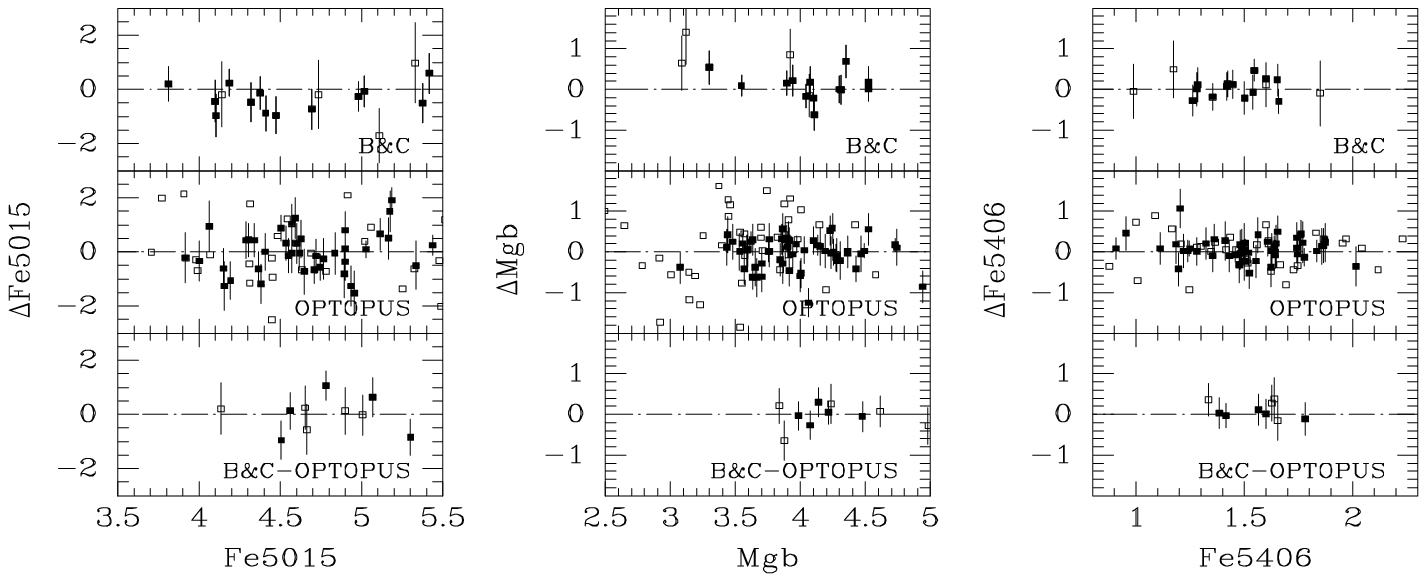}

\caption[ ]{Internal comparison.
Solid points -- data from spectra with S/N$\ge 20$ per {\AA}.
Open points -- data from spectra with S/N$< 20$ per {\AA}.
\label{fig-compint} }
\end{figure*}

Figure \ref{fig-compint} and Table \ref{tab-compint}
summarize the internal comparisons of the line indices.
Small offsets were applied to the OPTOPUS data for some indices as
follows.
Mgb$_{\rm B\&C}$ = Mgb$_{\rm OPTOPUS}+0.30$,
Fe5015$_{\rm B\&C}$ = Fe5015$_{\rm OPTOPUS}+0.31$,
Fe5270$_{\rm B\&C}$ = Fe5270$_{\rm OPTOPUS}+0.28$,
Fe5335$_{\rm B\&C}$ = Fe5335$_{\rm OPTOPUS}+0.13$, and 
Fe5406$_{\rm B\&C}$ = Fe5406$_{\rm OPTOPUS}+0.12$.
These offsets may be caused by poorer flux calibration of the 
OPTOPUS data.

The rms scatter of the comparisons is in general fully explained
by the estimated internal uncertainties, cf.\ Table \ref{tab-index}.
It is clear from the comparisons that the new index $\HbG$ has
lower uncertainty than the Lick/IDS index for $\Hb$.

\subsection{Transformation to Lick/IDS system}

The consistency with and transformation to the Lick/IDS system
were established by comparison with data from the Lick/IDS
project (Faber 1994) and from Gonz\'{a}les (1993).

Values from Faber (1994) were aperture corrected with our adopted
aperture correction under the assumption that they were all 
taken with or corrected to the Lick aperture size 
$1\farcs 5\times 4\farcs 0$ (equivalent to a circular aperture
with dia\-meter $2\farcs 95$, cf.\ Paper I).
Values from Gonz\'{a}les (1993) were aperture corrected.
The aperture size used by Gonz\'{a}les is
$2\farcs 1\times 5\farcs 0$, equivalent to a circular 
aperture with diameter $3\farcs 75$.
The comparisons are done for line indices uncorrected for the
velocity dispersion.

\begin{figure*}
\epsfxsize=17.8cm
\epsfbox{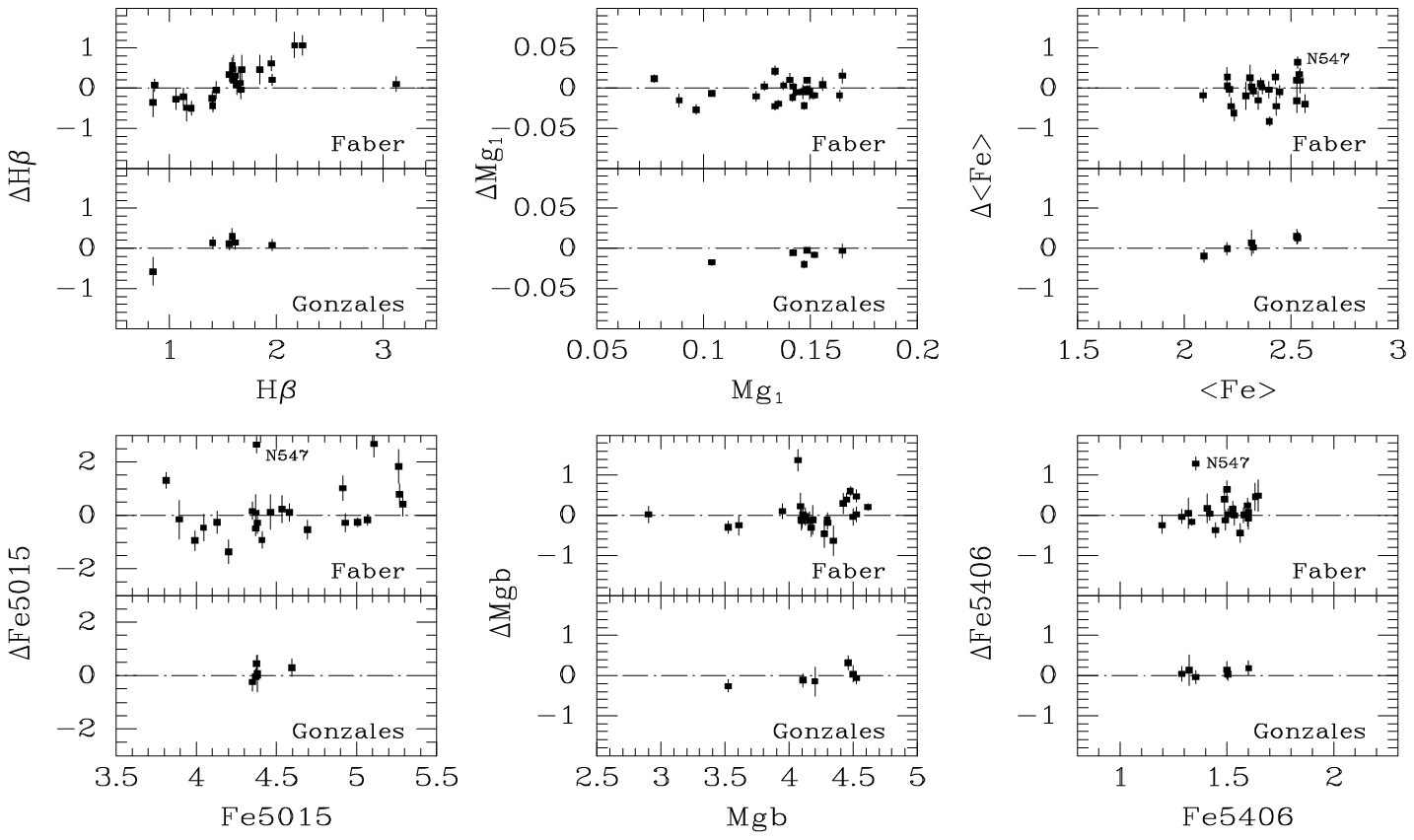}
\caption[ ]{External comparison.
All differences are calculated as ``our''-``literature'' and
plotted versus our determinations.
\label{fig-compext} }
\end{figure*}

The comparisons for some of the indices
are shown in Figure \ref{fig-compext}.
Table \ref{tab-compext} summarizes all the comparisons.
We have 24 galaxies in common with Faber and only 6 galaxies in
common with Gonz\'{a}les. It is however clear that the scatter
for the comparisons with data from Gonz\'{a}les is significantly lower
than for the comparisons with data from Faber.
As also noted by Worthey et al.\ (1992) the data from Gonz\'{a}les
are of significantly better quality than the Lick/IDS data.
The data from Gonz\'{a}les are also of sig\-ni\-fi\-cant\-ly better 
quality than the 
data presented in this paper, due to the very high signal-to-noise
spectra obtained by  Gonz\'{a}les.
The scatter in the comparisons with Gonz\'{a}les can be 
explained entirely by the internal uncertainty of our data.
The galaxy NGC547 is in both comparisons.
Our data agree with data from Gonz\'{a}les, while the data
from Faber show large deviations in some of the indices, 
see Figure \ref{fig-compext}.

The only significant offsets between our data and the Lick/IDS system
are the offset in $\Mgtwo$, consistent with the offset $-0.011$ adopted 
in Paper I, and the offset in $\Mgone$. 
In order to calibrate $\Mgone$ to the Lick/IDS system we add 
0.007 to our values.  The data given in Table \ref{tab-data1} 
are offset to the Lick/IDS system.

\begin{table}
\epsfxsize=9.5cm
\epsfbox{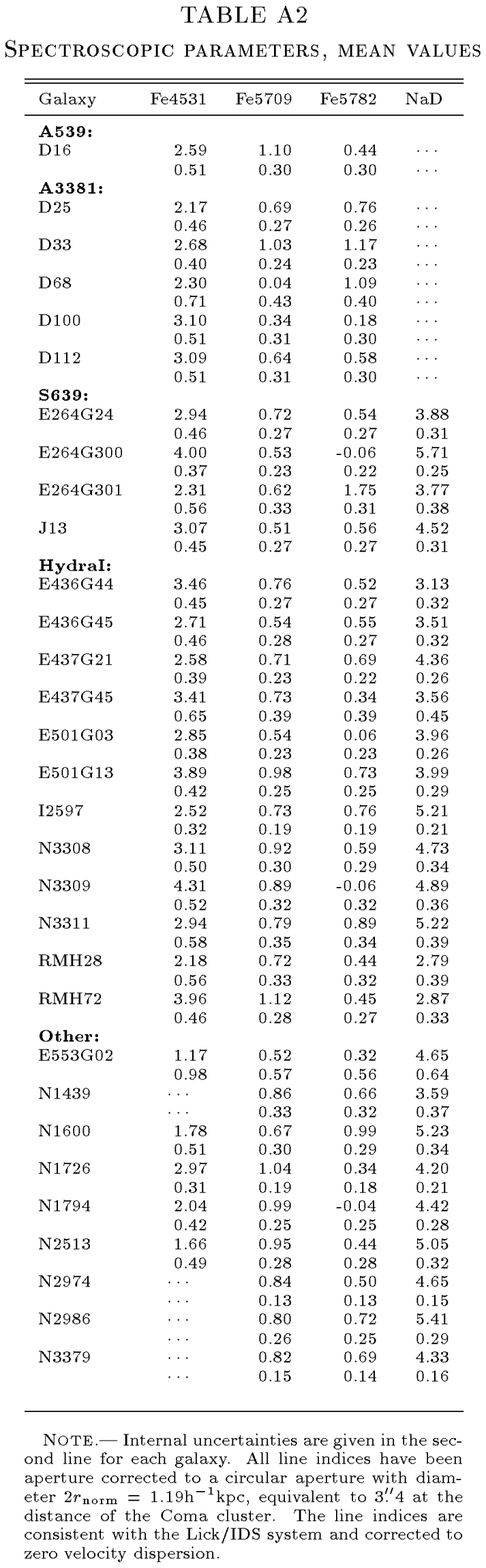}
\end{table}

\clearpage

\clearpage

\begin{table*}
\begin{minipage}{13cm}
\caption{Internal comparison \label{tab-compint} }
\begin{tabular}{lrrrrrrrrrr}
Index & \multicolumn{2}{c}{B\&C} & \multicolumn{2}{c}{B\&C$^a$} & 
\multicolumn{2}{c}{OPTOPUS} & \multicolumn{2}{c}{OPTOPUS$^a$} & 
\multicolumn{2}{c}{B\&C-OPTOPUS} \\
        & N$_{\rm gal}$ & rms & N$_{\rm gal}$ & rms &
N$_{\rm gal}$ & rms & N$_{\rm gal}$ & rms &
N$_{\rm gal}$ & rms \\ \hline
C4668   &  5 & 1.48 &  3 & 0.50 &    &      &    &      &    & \\
$\Hb$   & 17 & 0.63 & 13 & 0.31 & 11 & 0.60 &  4 & 0.69 &    & \\
$\HbG$  & 17 & 0.45 & 13 & 0.23 &  9 & 0.33 &  2 & 0.24 &    & \\
$\Mgone$& 17 & 0.011 &13 & 0.009 &56 & 0.017& 32 & 0.010& 10 & 0.29 \\
$\Mgtwo$& 17 & 0.019 &13 & 0.008 &56 & 0.021& 32 & 0.014&  3 & 0.009\\
Mgb     & 17 & 0.48 & 13 & 0.33 & 98 & 0.62 & 56 & 0.37 &  3 & 0.007\\
$\Fe$   & 17 & 0.35 & 13 & 0.19 & 97 & 0.43 & 56 & 0.30 & 10 & 0.35 \\
Fe5406  & 17 & 0.24 & 13 & 0.23 & 84 & 0.35 & 51 & 0.27 &  9 & 0.20 \\ \hline
\end{tabular}

Note -- $^a$ observations with S/N$<$20 per {\AA} excluded.
\end{minipage}
\end{table*}

\begin{table*}
\begin{minipage}{13cm}
\caption{External comparison \label{tab-compext} }
\begin{tabular}{lrrrrrrrrr}
Index & \multicolumn{3}{c}{Faber (1994)} & \multicolumn{3}{c}{Faber (1994)$^a$} & \multicolumn{3}{c}{Gonz\'{a}les (1993)} \\ 
& $N_{\rm gal}$ & $<\Delta >$ & rms 
& $N_{\rm gal}$ & $<\Delta >$ & rms 
& $N_{\rm gal}$ & $<\Delta >$ & rms \\ \hline
Fe4531 &    5 & -0.57 & 0.94 &  4 & -0.34 & 0.91 &   &      &  \\
C4668  &   15 &  0.49 & 1.90 & 12 &  0.07 & 1.44 &   &      &  \\
$\Hb$ &    24 &  0.15 & 0.44 & 21 &  0.09 & 0.34 & 6 & 0.03 & 0.31 \\
Fe5015 &   24 &  0.22 & 1.04 & 21 & -0.09 & 0.64 & 6 & 0.09 & 0.25 \\
$\Mgone$ & 24 & -0.004 & 0.013 & 21 & -0.006 & 0.011 & 6 & -0.009 & 0.007 \\
$\Mgtwo$ & 24 & -0.008 & 0.017 & 21 & -0.009 & 0.012 & 6 & -0.018 & 0.010 \\
Mgb &      24 &  0.04 & 0.40 & 21 & -0.02 & 0.27 & 6 & -0.04 & 0.20 \\
$\Fe$ &    24 & -0.07 & 0.34 & 21 & -0.09 & 0.31 & 6 &  0.09 & 0.18 \\
Fe5406 &   22 &  0.12 & 0.37 & 19 &  0.06 & 0.29 & 6 &  0.08 & 0.08 \\
Fe5709 &    7 &  0.08 & 0.18 &  5 &  0.13 & 0.17 &   &  &  \\
Fe5782 &    7 & -0.22 & 0.35 &  5 & -0.12 & 0.28 &   &  &  \\
NaD &       9 & -0.40 & 0.68 &  5 & -0.12 & 0.22 &   &  &  \\ \hline
\end{tabular}

Note -- Differences $<\Delta >$ are calculated as 
``our''-``literature''. \\
$^a$ observations with S/N$<$20 per {\AA} excluded.
\end{minipage}
\end{table*}

\clearpage

\subsection{Correlations between the magnesium indices}

\begin{figure}
\epsfxsize=8.5cm
\epsfbox{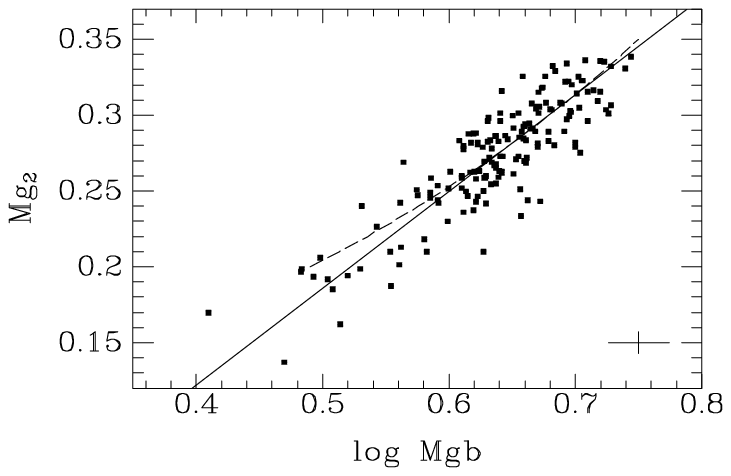}

\caption[ ]{The relation between the $\Mgtwo$ index and the Mgb index.
Solid line -- the relation given in Eq.\ \ref{eq-Mg}.
Dashed line -- the relation for E galaxies shown
by Burstein et al.\ (1984).
\label{fig-transMg} }
\end{figure}

The indices Mgb and $\Mgtwo$ are strongly correlated.
For the 161 E and S0 galaxies with both indices available we find
\begin{equation}
\begin{array}{ll}
\label{eq-Mg}
\Mgtwo = & \hspace*{7.5pt}0.638 \log\,{\rm Mgb} - 0.133 \\
         & \pm 0.044
\end{array}
\end{equation}
with an rms scatter of 0.019, see Figure \ref{fig-transMg}.
For $\Mgtwo$ in the interval 0.2--0.35 the relation is in
agreement with the relation for E galaxies shown by 
Burstein et al.\ (1984).
This relation is overplotted on Figure \ref{fig-transMg}.
Two galaxies with uncertainty on log\,Mgb larger than 0.08 were
excluded from the fit.
The relation has no significant intrinsic scatter.
The relation was therefore used to derive $\Mgtwo$ from
the measured Mgb for those 37 galaxies where $\Mgtwo$ could not
be measured because parts of the continuum bands for that index
were outside the observed wavelength range.

There is, as expected, a strong correlation between the indices 
$\Mgone$ and $\Mgtwo$. We find the following relation for E and S0
galaxies
\begin{equation}
\begin{array}{ll}
\Mgtwo = & \hspace*{7.5pt}1.515 \Mgone + 0.082 \\
         & \pm 0.050
\end{array}
\end{equation}
with an rms scatter of 0.015. 
The derived relation is very similar to the relation for E galaxies
shown by Burstein et al.\ (1984).
The relation has no significant intrinsic scatter. 
$\Mgone$ is not used in the present analysis, since this index from an
observational point of view does not contain additional information.

\end{document}